\documentstyle[12pt,titlepage,epsf]{article}
\begin{document}
\pagestyle{empty}
\newcommand{\gapproxeq}{\lower
.7ex\hbox{$\;\stackrel{\textstyle >}{\sim}\;$}}
\baselineskip=0.212in

\begin{flushleft}
\large
{SAGA-HE-81-95
\hfill August 4, 1995}  \\
\end{flushleft}

\vspace{0.7cm}

\begin{center}

\Large{{\bf Numerical solution of $\bf Q^2$ evolution equations}} \\
\vspace{0.3cm}
\Large{{\bf in a brute-force method}} \\

\vspace{0.7cm}

\Large
{M. Miyama and S. Kumano $^*$ }         \\

\vspace{0.5cm}

\Large
{Department of Physics, Saga University, Saga 840, Japan} \\


\Large{and }    \\


\Large
{Physics Department, Brookhaven National Laboratory} \\


{Upton, New York, 11973-5000, U.S.A.}         \\

\vspace{0.5cm}

\Large{ABSTRACT}

\end{center}

We investigate numerical solution of $Q^2$ evolution equations
for structure functions in the nucleon and in nuclei.
(Dokshitzer-Gribov-Lipatov-)Altarelli-Parisi
and Mueller-Qiu evolution equations are solved
in a brute-force method. Spin-independent flavor-nonsinglet and singlet
equations with next-to-leading-order $\alpha_s$ corrections
are studied. Dividing the variables $x$ and $Q^2$
into small steps, we simply solve the integrodifferential equations.
Numerical results indicate that accuracy is better than 2\%
in the region $10^{-4}<x<0.8$ if more than two-hundred $Q^2$ steps
and more than one-thousand $x$ steps are taken.
The numerical solution is discussed in detail, and
evolution results are compared with $Q^2$ dependent data
in CDHSW, SLAC, BCDMS, EMC, NMC, Fermilab-E665, ZEUS, and H1
experiments.
We provide a FORTRAN program for Q$^2$ evolution
(and ``devolution'') of nonsinglet-quark,
singlet-quark, $q_i+\bar q_i$, and gluon distributions
(and corresponding structure functions) in the nucleon and in nuclei.
This is a very useful program for studying spin-independent
structure functions.

\vspace{0.3cm}

\vfill

\noindent
{\rule{6.cm}{0.1mm}} \\

\vspace{-0.3cm}
\normalsize
\noindent
{* Email: 94sm10 or kumanos@cc.saga-u.ac.jp.
   Information on their research is available}  \\

\vspace{-0.6cm}
\noindent
{at http://www.cc.saga-u.ac.jp/saga-u/riko/physics/quantum1/structure.html
or} \\

\vspace{-0.6cm}
\noindent
\normalsize
{at ftp://ftp.cc.saga-u.ac.jp/pub/paper/riko/quantum1.} \\

\vspace{-0.2cm}
\hfill
{submitted for publication}

\vfill\eject
\pagestyle{plain}

\noindent
{\Large\bf {Program Summary}}
\vspace{0.4cm}

\medskip
\noindent
{\it Title of program:} BF1

\medskip
\noindent
{\it Computer:} AlphaServer 2100 4/200;
               {\it Installation:}
                     The Research Center for Nuclear Physics in Osaka

\medskip
\noindent
{\it Operating system:} OpenVMS V6.1

\medskip
\noindent
{\it Programming language used:} FORTRAN 77

\medskip
\noindent
{\it Peripherals used:} Laser printer

\medskip
\noindent
{\it No. of lines in distributed program, including test data, etc.:}
2439


\medskip
\noindent
{\it Nature of physical problem}

\noindent
This program solves Altarelli-Parisi equations
or modified evolution equations (Mueller-Qiu)
with or without next-to-leading-order $\alpha_s$ effects
for a spin-independent structure function or quark distribution.
Both flavor-nonsinglet and singlet cases are provided, so that
the distributions,
$xq_{_{NS}}$, $xq_{_S}$, $xq_i^+\equiv xq_i+x\bar q_i$ ($i$=quark flavor),
$xg$, $xF_{_{NS}}$, $xF_{_S}$, and $xF_i^+$
in the nucleon and in nuclei can be evolved.

\medskip
\noindent
{\it Method of solution}

\noindent
We divide the variable $x$ (and $Q^2$)
into very small steps, and
integration and differentiation are defined by
$\displaystyle{{{df(x)}\over{dx}}=
{{[f(x_{m+1})-f(x_m)]}\over{\Delta x_m}}}$ and
$\displaystyle{\int dx f(x)=\sum_{m=1}^{N_x} \Delta x_m f(x_m)}$.
Then, the integrodifferential equations are simply solved
step by step, and this method is so called brute-force method.
If the step numbers are increased, accurate results should
be obtained.

\medskip
\noindent
{\it Restrictions of the program}

\noindent
This program is used for calculating Q$^2$ evolution of a spin-independent
flavor-nonsinglet-quark, singlet-quark, $q_i^+$, and gluon distributions
(and corresponding structure functions)
in the leading order or in the next-to-leading-order of $\alpha_s$.
$Q^2$ evolution equations are the Altarelli-Parisi equations
and the modified ones (Mueller-Qiu).
The double precision arithmetic is used.
The renormalization scheme is the modified minimal subtraction
scheme ($\overline{MS}$).
A user provides the initial structure function or
quark distribution as a subroutine or as a data file.
Examples are explained in sections 4.2 and 4.3.
Then, the user inputs twenty-one parameters in section 4.1.

\medskip
\noindent
{\it Typical running time}

\noindent
Approximately five minutes on AlphaServer 2100 4/200 in the nonsinglet
case, sixty minutes in the singlet-quark evolution,
and eighty minutes in the singlet evolution with recombination effects.

\vfill\eject
\vspace{1.5cm}
\noindent
{\Large\bf {LONG WRITE-UP}}

\vspace{1.0cm}
\noindent
{\Large\bf {1. Introduction}}
\vspace{0.4cm}

Subnucleon degrees of freedom could be investigated
in high-energy lepton-nucleon interactions.
Measured structure functions depend in general on
two variables, $Q^2=-q^2$ and $x=Q^2/2P\cdot q$,
where $q$ is the four-momentum transfer and P is the nucleon momentum.
Bjorken scaling hypothesis suggests that
structure functions are independent of $Q^2$.
However, the scaling is not exactly satisfied and
they have weak logarithmic $Q^2$ dependence, which
is refer to as scaling violation.
Although the structure functions themselves cannot be calculated
except for lattice QCD methods, it is possible to estimate their $Q^2$
variations within perturbative QCD.

An intuitive way of describing the scaling violation is
to use the (Dokshitzer-Gribov-Lipatov-)Altarelli-Parisi equation \cite{AP}.
For example, the flavor-nonsinglet Altarelli-Parisi
(or DGLAP) equation is given by
$$
{\partial \over {\partial\ln Q^2}} \ q_{_{NS}}(x,Q^2)\ =
{{\alpha_s (Q^2)} \over {2\pi}} \ \int_x^1{dy\over y} \
P_{_{NS}}\biggl({x\over y}\biggr) \
q_{_{NS}}(y,Q^2) \ \ \ ,
\eqno{(1.1)}
$$
where $q_{_{NS}}(x,Q^2)$ is a nonsinglet quark distribution,
$P_{_{NS}}(x)$ is a nonsinglet splitting function, and
$\alpha_s(Q^2)$ is the running coupling constant.
The integrodifferential equation describes the progress that
a quark with the nucleon's momentum fraction $y$ radiates
a gluon and it becomes a quark with the momentum fraction $x$.
The splitting function determines
the probability of the splitting process.
The singlet Altarelli-Parisi equations are more complicated
due to gluon participation in the evolution process.
The flavor-singlet part is given by
coupled integrodifferential equations, which
are discussed in section 2.

We study not only the Altarelli-Parisi equations but also
modified evolution equations due to parton recombinations
(Mueller and Qiu) \cite{MQ}.
The recombination mechanism produces additional
terms in the Altarelli-Parisi evolution equations.
Because gluons play a major role at small $x$,
gluon recombination processes are investigated in Ref. \cite{MQ}.
Although the nonsinglet equation remains unchanged, the singlet
equations are modified due to the gluon recombinations.
The additional terms are nonlinear, so that it is not obvious
how to solve the evolution equation numerically.
On the other hand, the parton recombination mechanism is increasingly
important due to recent HERA data in the small $x$ region.
It is interesting to test whether the $F_2$ data at small $x$
could be related to higher-twist effects such as the recombination
contributions  \cite{GKR}.
In addition, the recombination mechanism is used for explaining
nuclear shadowing \cite{SKF2}, and it is an important factor
in $Q^2$ evolution of nuclear structure functions.

The $Q^2$ evolution equations are often used in experimental analysis
and also in theoretical calculations, so that it is worth while
creating a computer program in solving the equations accurately.
A number of methods have been developed \cite{SOLVE}, and
they include brute-force methods, Mellin-transformation methods,
and orthogonal polynomial methods.
Among these methods, the Laguerre-polynomial one \cite{LAG}
is considered an efficient one in computing time and
in numerical accuracy.
We have been investigating the Laguerre method and find that
accuracy may not be very satisfactory at small $x$ and at large $x$
particularly in the nonsinglet case.
With development of high-energy accelerators, the small $x$ region
becomes important. In fact, the $x$ region of $10^{-4}$ could be
reached with large enough $Q^2$ at which perturbative QCD could be used.
Considering structure-function studies at HERA and at future
high-energy laboratories, we should have a computer program
which enables the evolution with good accuracy at small $x$.
In addition to this problem,
there is other difficulty in the Laguerre method.
It is not obvious how to apply the Laguerre method
to the case of modified $Q^2$ evolution equations with parton recombinations.
It is because of the existence of nonlinear recombination terms
in the evolution equations.

These issues motivated us to explore an alternative method.
As a possible way to solve the above difficulties, we study
a brute-force method.
This method was, for example, investigated in Ref. \cite{BF};
however, there is no published article in discussing the details
of the solution and its accuracy.
The method is perhaps the simplest one in solving the
integrodifferential equations.
We divide the variables $x$ and $Q^2$ into very small steps,
and integration and differentiation are defined by
$df(x)/dx=[f(x_{m+1})-f(x_m)]/\Delta x_m$ and
$\displaystyle{\int dx f(x)=\sum_{m=1}^{N_x} \Delta x_m f(x_m)}$.
Then, the evolution equation can be solved step by step.
If $N_x$ is large enough, we should be able to get accurate
numerical results. Our research purposes are
to investigate the details of numerical accuracy
and to provide a useful computer program in the brute-force method.

The evolution equations, which are solved by the brute-force method,
are given in section 2.
We explain the brute-force method in section 3.
In section 4, information for running the program BF1 is supplied.
Each subroutine in the program is explained in section 5.
Numerical results and comparisons with experimental data
are given in section 6.
The results are summarized in section 7.
Explicit equations of the splitting functions and
other necessary quantities are given in appendices.

\vfill\eject
\vspace{1.0cm}
\noindent
{\Large\bf {2. $\bf Q^2$ evolution equations}}
\vspace{0.4cm}

We solve two types of evolution equations.
One is the ordinary equations, so called Altarelli and Parisi equations,
and the other is the modified ones due to recombinations by Mueller and Qiu.
Parton distributions or structure functions in the nucleon can be
evolved in both methods. The modified equations can also handle evolution
of nuclear parton distributions.

\vspace{0.8cm}
\noindent
{\it 2.1 (Dokshitzer-Gribov-Lipatov-)Altarelli-Parisi equations}
\vspace{0.4cm}

The nonsinglet ``Altarelli-Parisi'' equation is given in Eq. (1.1),
which is valid both in the leading order (LO) case
and in the next-to-leading order (NLO) one.
NLO effects can be included in the running coupling constant
$\alpha_s(t)$ and in the splitting functions $P_{ij}(z)$.
In order to remove the extra $Q^2$ dependence in front of
the integral in Eq. (1.1), we use the variable $t$ defined by
$$
t \equiv -{2\over \beta_0}\ln
\biggl[{\alpha_s(Q^2)\over \alpha_s(Q_0^2)}\biggr]   \ \ \ ,
\eqno{(2.1)}
$$
instead of $Q^2$.
The parton distribution and the splitting function
multiplied by $x$ satisfy the same integrodifferential equation.
Therefore, defining $\widetilde f(x)$ by
$$
\widetilde f (x)=x f(x) \ \ \ ,
\eqno{(2.2)}
$$
we rewrite the evolution equation as
$$
{\partial \over {\partial t}} \ \widetilde q_{_{NS}}(x,t)\
= \ \int_x^1{dy\over y} \
\widetilde P_{_{NS}}\biggl({x\over y}\biggr) \
\widetilde q_{_{NS}}(y,t) \ \ \ .
\eqno{(2.3)}
$$

The singlet case is more complicated than the nonsinglet one
due to the gluon participation in the evolution,
and it becomes coupled integrodifferential equations.
The singlet quark distribution is defined by
$\displaystyle{q_{_S}(x,t)= \sum_i q_i^+(x,t)}$,
where $i$ is the quark flavor and the $q_i^+$ distribution is given by
$q_i^+(x,t)=q_i(x,t)+\bar q_i(x,t)$.
We write evolution equations in terms of the $q_i^+$ distribution
\cite{HWW,OTHERAP}:
$$
{\partial \over {\partial t}} \ \widetilde {q}_i^{\ +}\left({x,t}\right)\ =\
\int_{x}^{1}{dy \over y}\
\left[{\ \sum_j \widetilde {P}_{q_{i}^+ q_{j}^+}\left({{x \over y}}\right)\
\widetilde q_j^{\ +} \left({y,t}\right)\
+\  \widetilde {P}_{qg}\left({{x \over y}}\right)\
\widetilde g\left({y,t}\right)\ }\right]
\ \ \ ,
\eqno{(2.4a)}
$$
$$
{\partial \over {\partial t}} \ \widetilde g\left({x,t}\right)\ =\
\int_{x}^{1}{dy \over y}\
\left[{\ \sum_j \widetilde {P}_{gq_j^+}\left({{x \over y}}\right)\
\widetilde q_j^{\ +}\left({y,t}\right)\
+\  \widetilde {P}_{gg}\left({{x \over y}}\right)\
\widetilde g\left({y,t}\right)\ }\right]
\ \ \ .
\eqno{(2.4b)}
$$
Each term in Eqs. (2.4a) and (2.4b) describes the process
that a parton $p_j$ with the nucleon's momentum fraction $y$
splits into a parton $p_i$ with the momentum fraction $x$
and another parton.
The splitting function $P_{p_i p_j}(z)$ determines
the probability that such a splitting process occurs and
the $p_j$-parton momentum is reduced by the fraction $z$.
Using $\widetilde {P}_{q_{i}^+ q_{j}^+}^{(1)}(x)
=$$\delta_{ij} \widetilde P_{q^+,NS}^{(1)}$$+2C_{_F}T_R x F_{qq}(x)$
\cite{HWW}
and $\widetilde {P}_{gq_j^+}(x)=$ (independent of $j$), we obtain
$$
{\partial \over {\partial t}} \ \widetilde {q}_i^{\ +}\left({x,t}\right)\ =\
\int_{x}^{1}{dy \over y}\
\left[\
\widetilde {P}_{q^+,\  NS}\left({{x \over y}}\right)\
\widetilde q_i^{\ +} \left({y,t}\right)\
+\ 2 C_{_F} T_R \widetilde F_{qq} \left({{x \over y}}\right)\
\widetilde q_{_S} \left({y,t}\right)\
\right.
$$
$$
\left.
\ \ \ \ \ \ \ \ \ \ \ \ \ \ \ \ \ \ \ \ \ \ \ \ \ \ \ \
\ \ \ \ \ \ \ \ \ \ \ \ \ \ \ \ \ \ \ \ \ \ \ \ \ \ \ \
\ \ \ \ \ \ \ \ \ \
+\  \widetilde {P}_{qg}\left({{x \over y}}\right)\
\widetilde g\left({y,t}\right)\ \right]
\ \ \ ,
\eqno{(2.5a)}
$$
$$
{\partial \over {\partial t}} \ \widetilde g\left({x,t}\right)\ =\
\int_{x}^{1}{dy \over y}\
\left[{\ \widetilde {P}_{gq^+}\left({{x \over y}}\right)\
\widetilde q_{_S} \left({y,t}\right)\
+\  \widetilde {P}_{gg}\left({{x \over y}}\right)\
\widetilde g\left({y,t}\right)\ }\right]
\ \ \ .
\eqno{(2.5b)}
$$
All the splitting functions in Eqs. (2.5a) and (2.5b) are
listed in Appendix B.
If the summation over $i$ is taken in Eq. (2.5a), it becomes
the evolution equation for the singlet distribution $q_{_S}$:
$$
{\partial \over {\partial t}} \ \widetilde {q}_{_S}\left({x,t}\right)\ =\
\int_{x}^{1}{dy \over y}\
\left[\
\left\{\ \widetilde {P}_{q^+,\ NS}\left({{x \over y}}\right)\
+\ 2 N_f C_{_F} T_R \widetilde F_{qq} \left({{x \over y}}\right)\ \right\}\
\widetilde q_{_S} \left({y,t}\right)\
\right.
$$
$$
\left.
\ \ \ \ \ \ \ \ \ \ \ \ \ \ \ \ \ \ \ \ \ \ \ \ \ \ \ \
\ \ \ \ \ \ \ \ \ \ \ \ \ \ \ \ \ \ \ \ \ \ \ \ \ \ \ \
\ \ \ \ \ \ \ \ \ \
+\  N_f \widetilde {P}_{qg}\left({{x \over y}}\right)\
\widetilde g\left({y,t}\right)\ \right]
\ \ \ .
\eqno{(2.6)}
$$
The reason for writing the evolution equations in term of
the flavor-dependent distribution $q_i^+$ instead of
the singlet one $q_{_S}$ is because antiquark distributions
are neither SU(3)$_{\rm flavor}$ nor SU(2)$_{\rm flavor}$ symmetric.

Next, we discuss NLO effects in the evolution equations.
NLO effects are included in the coupling constant $\alpha_s(t)$,
in the splitting functions $P_{p_i p_j}(z)$, and
in coefficient functions.
Explicit NLO expressions are given in appendices.
It should be noted that the LO or NLO expression
of $\alpha_s(Q^2)$ is used in the LO or NLO evolution case respectively
in Eq. (2.1).
Once the NLO corrections are included in the evolution,
the renormalization scheme has to be specified.
Throughout this paper, we use the $\overline {MS}$ scheme.
The splitting function is given by the LO and NLO ones as
$$
P_{ij}(x)=P_{ij}^{(0)}(x)
           +{{\alpha_s(Q^2)} \over{2\pi}} P_{ij}^{(1)}(x) \ \ \ .
\eqno{(2.7)}
$$
Then the splitting functions $\widetilde P_{ij}(z)$
in Eqs. (2.4a) and (2.4b) are
$$
\widetilde P_{ij}(x) = \widetilde P_{ij}^{(0)}(x)
                +{{\alpha_s(t)} \over{2\pi}} R_{ij}(x)
\ \ \ ,
\eqno{(2.8)}
$$
where the function $R_{ij}(z)$ is
$$
R_{ij} (x)\equiv \widetilde{P}_{ij}^{(1)}(x)-{\beta_1\over 2\beta_0}
\widetilde{P}_{ij}^{(0)}(x)  \ \ \ .
\eqno{(2.9)}
$$
The second term in Eq. (2.9)
appears because of the transformation from $Q^2$ to $t$.
To be precise, the splitting functions $\widetilde P_{ij}$
in Eqs. (2.3)--(2.6) and (2.8) should be denoted,
for example, $\widetilde P_{ij}'$ because
it is different from
$xP_{ij}=xP_{ij}^{(0)}$$+(\alpha_s/2\pi)xP_{ij}^{(1)}$.
However, we omit the prime throughout this paper for using
simpler notations.

We discussed the evolution equations for
quark and gluon distributions.
In calculating structure functions, a correction due to the NLO
coefficient function should be taken into account.
In the nonsinglet case, it is given by
$$
\widetilde{F}_{n}^{NS}(x,{Q}^{2})\ =\
\int_{x}^{1}{dy \over y}\
\widetilde{C}_{n}^{q}\left({{x \over y},{\alpha }_{s}}\right)\
\widetilde{q}_{n}^{NS}(y,{Q}^{2})
\ \ \ ,
\eqno{(2.10)}
$$
where $n$ denotes the type of a structure function,
$\tilde F_n$=$xF_1$, $F_2$, or $xF_3$ for $n$=1, 2, or 3 respectively,
and  ${C}_{n}^{q}$ is a quark coefficient function.
In the singlet case or in the $q_i^+$ case,
an additional gluon correction term
should be taken into account:
$$
\widetilde{F}_{n, i}^+ (x,{Q}^{2})\ =\
\int_{x}^{1}{dy \over y}\
\widetilde{C}_{n}^{q}\left({{x \over y},{\alpha }_{s}}\right)\
\widetilde{q}_{i}^+ (y,{Q}^{2})
\ + \
\int_{x}^{1}{dy \over y}\
\widetilde{C}_n^{g}\left({{x \over y},{\alpha }_{s}}\right)\
\widetilde{g}(y,{Q}^{2})
\ \ \ ,
\eqno{(2.11)}
$$
where ${C}_n^{g}$ is a gluon coefficient function.
$C_n^q$ and $C_n^g$ are given in Appendix C.
The $Q^2$ evolution of a quark distribution is first calculated, then
the structure function at $Q^2$ is evaluated by using the convolution
integrals in Eq. (2.10) or (2.11), so that this method is called
``two-step evolution''.

The above procedure requires an initial quark (and gluon) distribution
for getting a structure function at certain $Q^2$.
It cannot be used for the NLO evolution if a structure function
is given as the initial distribution.
In such a case, ``one-step-evolution'' equations for
the structure function are useful. They are derived from the evolution
equations and the convolution equations with the coefficient functions.
The nonsinglet equation is in the same form with Eqs. (2.3) and (2.8)
\cite{LAG}:
$$
{\partial \over {\partial t}}
\ \widetilde F_{_{NS}}(x,t)\ = \ \int_x^1{dy\over y} \
\biggl [ \
\widetilde P_{_{NS}}^{(0)}\biggl({x\over y}\biggr)
        +{{\alpha_s(t)} \over{2\pi}} R_{_{NS}}\biggl({x\over y}\biggr) \
\biggr ] \
\widetilde F_{_{NS}}(y,t) \ \ \ ,
\eqno{(2.12)}
$$
except for taking
$$
R_{_{NS}} (x) = \widetilde{P}_{_{NS}}^{(1)}(x)
-{\beta_1\over 2\beta_0}
\widetilde{P}_{_{NS}}^{(0)}(x)
- {{\beta_0} \over 4} \widetilde B_n^{q} (x)
\ \ \ .
\eqno{(2.13)}
$$
$B_n^{q} (x)$ ($n$=1, 2, or 3)
is the $\alpha_s$ correction in the coefficient
function in Appendix C.
Using the above one-step equation, we can evolve the
nonsinglet structure function itself.
In the similar way, the one-step $F_i^+$ evolution equation
is given by \cite{HWW}
$$
{\partial \over {\partial t}}
\left({\begin{array}{c}{\widetilde {F}_i}^{\ +}\left({x,t}\right)\\
\widetilde {g}(x,t)\end{array}}\right)\
=\ \int_{x}^{1}{\frac{dy}{y}}\
\left[ {\widetilde {\bf P}}^{\left({0}\right)} \left(x \over y \right)
\ +\ {\frac{{\alpha }_{s}\left({t}\right)}{2\pi }}\
{\bf R} \left(x \over y \right) \right]
\ \left({\begin{array}{c}{\widetilde {F}_j}^{\ +}\left({y,t}\right)\\
\widetilde {g}(y,t)\end{array}}\right)
\ \ \ ,
\eqno{(2.14)}
$$
with
$$
{\bf R} (x) \ = \
{\widetilde {\bf P}}^{\left({1}\right)} (x)
\ -\ {\frac{{\beta }_{1}}{2{\beta }_{0}}}\
{\widetilde {\bf P}}^{\left({0}\right)} (x)
\ -\ {\frac{{\beta }_{0}}{4}}\
\widetilde {\bf D} ^{\left({1}\right)} (x)
\ +\ \widetilde {\bf E} (x)
\ \ \ .
\eqno{(2.15)}
$$
${\bf P}=\widetilde {\bf P}/x$,
${\bf D}^{(1)}=\widetilde {\bf D}^{(1)}/x$, and
${\bf E}=\widetilde {\bf E}/x$ are given by
$$
{\bf P}^{(0),(1)} \left({x}\right)\ =
\ \left({\begin{array}{cc}
{\displaystyle \sum_j} P_{q_i^+ q_j^+}^{(0),(1)}
\left({x}\right)&P_{qg}^{(0),(1)} \left({x}\right) \\
{\displaystyle \sum_j} P_{gq_j^+}^{(0),(1)}
\left({x}\right)&P_{gg}^{(0),(1)}
\left({x}\right)
\end{array}}\right)
\ \ \ ,
\eqno{(2.16)}
$$
$$
{\bf D}^{\left({1}\right)}\left({x}\right)\ = \rm
\ \left({\begin{array}{cc}B_n^q(x)&B_n^g(x) \\
0&0\end{array}}\right)
\ \ \ ,
\eqno{(2.17)}
$$
and
$$
{\bf E}\left({x}\right)\ = \rm
\ \left({\begin{array}{cc}{E}_{qq}
\left({x}\right)&{E}_{qg}\left({x}\right) \\
{E}_{gq}\left({x}\right)&{E}_{gg}
\left({x}\right)\end{array}}\right)
\ \ \ .
\eqno{(2.18)}
$$
Each matrix element in Eqs. (2.16), (2.17), and (2.18) is
given in Appendix B.
It should be mentioned that the above one-step-evolution equation
is slightly different from the one in Ref. \cite{HWW}.
This discrepancy is because we use
a convention for the momentum conservation,
$\displaystyle{\int_0^1 dx [\sum_i x (q_i + \bar q_i) + xg ] = 1}$,
which is different from the one used by Herrod, Wada, and Webber,
$\displaystyle{\int_0^1 dx [F_S + xg' ] = 1}$.
Therefore, the matrices ${\bf D} ^{(1)}$
and $\bf E$ are different from those listed in their paper.

\vspace{0.8cm}
\noindent
{\it 2.2 Modified evolution equations due to recombinations
         (Mueller-Qiu equations)}
\vspace{0.4cm}

The Altarelli-Parisi evolution equations have been
used extensively and they have been successful in
describing many experimental data.
However, as it becomes possible to reach the small $x$ region
by high-energy accelerators, it is necessary to investigate
the details of small $x$ physics.
The gluon distribution $g(x,Q^2)$ is a measurement of a gluon
with transverse size $\sim 1/\sqrt{Q^2}$, and
the number of gluons per unit rapidity is $xg(x,Q^2)$.
If $xg(x,Q^2)/Q^2 << R_0^2$ ($R_0$=nucleon size)
is satisfied, gluon interactions
are neglected and the Altarelli-Parisi equations in Eqs. (2.3), (2.4),
(2.5), and (2.6)
can be used for the $Q^2$ evolution.
However, if the transverse area $xg(x,Q^2)/Q^2$ exceed the nucleon size
$\sim R_0^2$, gluons within a unit of rapidity spatially overlap, and
the gluons are no longer considered as free partons.
If the gluon density becomes large $xg(x,Q^2) \gapproxeq R_0^2 Q^2$,
the gluon interactions should be taken into account.
These interactions are called parton recombinations, which are also
used for explaining nuclear shadowing \cite{SKF2}.
There are a number of studies on the recombinations.
Among them, we employ the evolution equations proposed by
Mueller and Qiu \cite{MQ}. They investigated gluon-gluon recombination
effects on the evolution, and they proposed
the following modified evolution equations:
\vfill\eject
$$
{\partial \over {\partial t}} \ \widetilde {q}_i^{\ +}\left({x,t}\right)\ =\
\int_{x}^{1}{dy \over y}\
\left[\
\widetilde {P}_{q^+,\  NS}\left({{x \over y}}\right)\
\widetilde q_i^{\ +} \left({y,t}\right)\
+\ 2C_{_F} T_R \widetilde F_{qq} \left({{x \over y}}\right)\
\widetilde q_{_S} \left({y,t}\right)\
\right.
$$
$$
\left.
\ \ \ \ \ \ \ \ \ \ \ \ \ \ \ \ \ \ \ \ \ \ \ \ \ \ \ \
\ \ \ \ \ \ \ \ \ \ \ \ \ \ \ \ \ \ \ \ \ \ \ \ \ \ \ \
\ \ \ \ \ \ \ \ \ \
+\  \widetilde {P}_{qg}\left({{x \over y}}\right)\
\widetilde g\left({y,t}\right)\ \right]
$$
$$
\ \ \ \ \ \ \ \ \ \ \
-\ 2\ {K \over \pi{R_0}^{2}}\ {1 \over {Q}^{2}}\
\left\{{{(2\pi)^2 {\alpha }_s (t)
\over {\rm N}_{\rm c}\left({{\rm N}_{\rm c}^{\rm 2}\rm -1}\right)}\
\left({{4 \over 15}{N}_{c}^{2}\ -\ {3 \over 5}}\right)\
{\left[{\widetilde g\left({x,t}\right)}\right]}^{2}}
\right.
$$
$$
\ \ \ \ \ \ \ \ \ \ \ \ \ \ \ \ \ \ \ \ \ \
\ \ \ \ \ \ \ \ \ \ \ \ \ \ \ \ \ \ \ \ \ \
\ \ \ \ \ \ \ \
\left.
+\ \int_{x}^{1}{dy \over y}\
\widetilde {\bar P}_{qg}\left({{x \over y}}\right)\
\widetilde {G}_{HT}\left({y,t}\right) \
\right\}
\ \ \ ,
\eqno{(2.19a)}
$$
$$
{\partial \over {\partial t}} \ \widetilde {q}_{_S}\left({x,t}\right)\ =\
\int_{x}^{1}{dy \over y}\
\left[\
\left\{\ \widetilde {P}_{q^+ NS} \left({{x \over y}}\right)\
+\ 2 N_f C_{_F} T_R \widetilde F_{qq} \left({{x \over y}}\right)\ \right\}\
\widetilde q_{_S} \left({y,t}\right)\
\right.
$$
$$
\left.
\ \ \ \ \ \ \ \ \ \ \ \ \ \ \ \ \ \ \ \ \ \ \ \ \ \ \ \
\ \ \ \ \ \ \ \ \ \ \ \ \ \ \ \ \ \ \ \ \ \ \ \ \ \ \ \
\ \ \ \ \ \ \ \ \ \
+\  N_f \widetilde {P}_{qg}\left({{x \over y}}\right)\
\widetilde g\left({y,t}\right)\ \right]
$$
$$
\ \ \ \ \ \ \ \ \ \ \
-\ 2 N_f\ {K \over \pi{R_0}^{2}}\ {1 \over {Q}^{2}}\
\left\{{{(2\pi)^2 {\alpha }_s (t)
\over {\rm N}_{\rm c}\left({{\rm N}_{\rm c}^{\rm 2}\rm -1}\right)}\
\left({{4 \over 15}{N}_{c}^{2}\ -\ {3 \over 5}}\right)\
{\left[{\widetilde g\left({x,t}\right)}\right]}^{2}}
\right.
$$
$$
\ \ \ \ \ \ \ \ \ \ \ \ \ \ \ \ \ \ \ \ \ \
\ \ \ \ \ \ \ \ \ \ \ \ \ \ \ \ \ \ \ \ \ \
\ \ \ \ \ \ \ \
\left.
+\ \int_{x}^{1}{dy \over y}\
\widetilde {\bar P}_{qg}\left({{x \over y}}\right)\
\widetilde {G}_{HT}\left({y,t}\right) \
\right\}
\ \ \ ,
\eqno{(2.19b)}
$$
$$
{\partial \over {\partial t}} \ \widetilde g\left({x,t}\right)\ =\
\int_{x}^{1}{dy \over y}\
\left[{\ \widetilde {P}_{gq^+}\left({{x \over y}}\right)\
\widetilde q_{_S}\left({y,t}\right)\
+\  \widetilde {P}_{gg}\left({{x \over y}}\right)\
\widetilde g\left({y,t}\right)\ }\right]
$$
$$
\ \ \ \ \ \ \ \ \ \ \ \ \ \ \ \ \ \ \ \ \ \
\ \ \ \ \ \ \ \ \
-\ {K \over \pi{R_0}^{2}}\ {1 \over {Q}^{2}}\
\left\{{{8{\pi }^{2} \over {N}_{c}^{2}-1} \
\alpha_s (t) \ {C}_{G}^2 \
\int_{x}^{1}{dy \over y}\
{\left[{\widetilde g\left({y,t}\right)}\right]}^{2}}\right\}
\ \ \ ,
\eqno{(2.19c)}
$$
where $\widetilde G_{HT}(x,t)$ is a higher-dimensional distribution, which
follows the evolution equation:
$$
{\partial \over {\partial t}}\
\widetilde {G}_{HT}\left({x,t}\right)\ =\
{8{\pi }^{2} \over {N}_{c}^{2}-1} \
\alpha_s(t)\ {C}_{G}^2\
\int_{x}^{1}{dy \over y}\
{\left[{\widetilde g\left({y,t}\right)}\right]}^{2}
\ \ \ .
\eqno{(2.19d)}
$$
In order to solve this equation, the initial $\widetilde G_{HT}$
distribution has to be supplied. Because $\widetilde G_{HT}$
is associated with the two gluon interactions, we may
assume
$$
\widetilde G_{HT}(x,Q_0^2)\ =\ K_{HT} [\widetilde g(x,Q_0^2)]^2
\ \ \ ,
\eqno{(2.20)}
$$
where $K_{HT}$ is a constant.
In Eq. (2.19a, b, c), $R_0$ is the nucleon size,
$N_f$ is the number of flavor, $N_c$ is the number of color,
and $K$=9/8 so that nuclear correction terms vanish
for $A=1$ [See Eqs. (2.22a, b, c)].
The splitting function $\widetilde {\bar P}_{qg} (z)$
is given in Appendix B.
The recombination contributions enter into the evolution
in the same $\alpha_s$ order as the next-to-leading order
in the splitting function, so that the NLO effects
in the Altarelli-Parisi evolution terms should be included.

We comment on additional terms in Eq. (2.19).
For the details the recombination processes,
we refer the reader to the original paper \cite{MQ}.
The recombination terms are related to
the two-gluon density,
$G^{(2)}(x,Q^2)$$=G^{(2)}(x_1,x_2,p_{1T}^2,p_{2T}^2)$ with
$x_1=x_2=x$ and $p_{1T}^2=p_{2T}^2=Q^2$.
Because the normalization of the two-gluon density is given by
$G^{(2)}(x_1,x_2,p_{1T}^2,p_{2T}^2)$$=(3R_0\bar n_0/2)
g(x_1,p_{1T}^2) g(x_2,p_{2T}^2)$ with
$\bar n_0=3/4\pi R_0^3$,
it becomes $G^{(2)}(x,Q^2)=(3R_0\bar n_0/2) [g(x,Q^2)]^2$,
which are the $ [g(x,Q^2)]^2$ terms in Eq. (2.19).
It results in the recombination factor, which is proportional
to $R_0 \bar n_0 \alpha_s/ Q^2= 3\alpha_s/(4 \pi R_0^2 Q^2)$.
The factor $\alpha_s/Q^2$ arises because a
parton-parton fusion cross section
is proportional to $\alpha_s/Q^2$.
This extra $1/Q^2$ dependence
is an interesting higher-twist effect,
which could be investigated experimentally at small $x$ and at small $Q^2$.
The $1/R_0^2$ factor arises due to the integration over
the intrinsic transverse momentum.

In the case of nuclear parton distributions, there are additional terms
due to parton recombinations from different nucleons.
Parton distributions in a nucleus
are defined by
$$
\widetilde p^A(x,t)=\widetilde p(x,t)-\delta \widetilde p^A(x,t) \ \ \ ,
\eqno{(2.21)}
$$
where nuclear distributions are expressed by those per nucleon.
The evolution of the first term is given in Eq. (19).
The evolution of the nuclear correction terms is obtained by solving
$$
{\partial \over {\partial t}} \
\delta \widetilde {q}_i^{\ +,A}\left({x,t}\right)\ =\
\int_{x}^{1}{dy \over y}\
\left[\
\widetilde {P}_{q^+,\  NS}\left({{x \over y}}\right)\
\delta \widetilde q_i^{\ +,A} \left({y,t}\right)\
+\ 2C_{_F} T_R \widetilde F_{qq} \left({{x \over y}}\right)\
\delta \widetilde q_{_S}^{^{A}} \left({y,t}\right)\
\right.
$$
$$
\left.
\ \ \ \ \ \ \ \ \ \ \ \ \ \ \ \ \ \ \ \ \ \ \ \ \ \ \ \
\ \ \ \ \ \ \ \ \ \ \ \ \ \ \ \ \ \ \ \ \ \ \ \ \ \ \ \
\ \ \ \ \ \ \ \ \ \
+\  \widetilde {P}_{qg}\left({{x \over y}}\right)\
\delta \widetilde g^{^{A}} \left({y,t}\right)\ \right]
\ \ \ ,
$$
$$
\ \ \ \ \ \ \ \ \ \ \
+\ 2
\left({3 \over 2}R_A\overline{n}\ -\ {K \over \pi {R_0}^{2}}\right)\
{1 \over {Q}^{2}}\ \left\{{{(2\pi)^2 {\alpha }_s (t)
\over {\rm N}_{\rm c}\left({{\rm N}_{\rm c}^{\rm 2}\rm -1}\right)}\
\left({{4 \over 15}{N}_{c}^{2}\ -\ {3 \over 5}}\right)\
{\left[{\widetilde g\left({x,t}\right)}\right]}^{2}}
\right.
$$
$$
\ \ \ \ \ \ \ \ \ \ \ \ \ \ \ \ \ \ \ \ \ \
\ \ \ \ \ \ \ \ \ \ \ \ \ \ \ \ \ \ \ \ \ \
\ \ \ \ \ \ \ \
\left.
+\ \int_{x}^{1}{dy \over y}\
\widetilde {\bar P}_{qg}\left({{x \over y}}\right)\
\widetilde {G}_{HT}\left({y,t}\right) \
\right\}
\ \ \ ,
\eqno{(2.22a)}
$$
$$
{\partial \over {\partial t}} \
\delta \widetilde {q}_{_S}^{^{\ A}} \left({x,t}\right)\ =\
\int_{x}^{1}{dy \over y}\
\left[\
\left\{\ \widetilde {P}_{q^+, NS}\left({{x \over y}}\right)\
+\ 2 N_f C_{_F} T_R \widetilde F_{qq} \left({{x \over y}}\right)\ \right\}\
\delta \widetilde q_{_S}^{^{A}} \left({y,t}\right)\
\right.
$$
$$
\left.
\ \ \ \ \ \ \ \ \ \ \ \ \ \ \ \ \ \ \ \ \ \ \ \ \ \ \ \
\ \ \ \ \ \ \ \ \ \ \ \ \ \ \ \ \ \ \ \ \ \ \ \ \ \ \ \
\ \ \ \ \ \ \ \ \ \
+\  N_f \widetilde {P}_{qg}\left({{x \over y}}\right)\
\delta \widetilde g^{^{A}} \left({y,t}\right)\ \right]
$$
$$
\ \ \ \ \ \ \ \ \ \ \
+\ 2 N_f
\left({3 \over 2}R_A\overline{n}\ -\ {K \over \pi {R_0}^{2}}\right)\
{1 \over {Q}^{2}}\ \left\{{{(2\pi)^2 {\alpha }_s (t)
\over {\rm N}_{\rm c}\left({{\rm N}_{\rm c}^{\rm 2}\rm -1}\right)}\
\left({{4 \over 15}{N}_{c}^{2}\ -\ {3 \over 5}}\right)\
{\left[{\widetilde g\left({x,t}\right)}\right]}^{2}}
\right.
$$
$$
\ \ \ \ \ \ \ \ \ \ \ \ \ \ \ \ \ \ \ \ \ \
\ \ \ \ \ \ \ \ \ \ \ \ \ \ \ \ \ \ \ \ \ \
\ \ \ \ \ \ \ \
\left.
+\ \int_{x}^{1}{dy \over y}\
\widetilde {\bar P}_{qg}\left({{x \over y}}\right)\
\widetilde {G}_{HT}\left({y,t}\right) \
\right\}
\ \ \ ,
\eqno{(2.22b)}
$$
\vfill\eject
$$
{\partial \over {\partial t}} \
\delta \widetilde {g}^{^{\ A}}\left({x,t}\right)\
=\ \int_{x}^{1}{dy \over y}\ \left[{\
\widetilde {P}_{g q^+}\left({{x \over y}}\right)\
\delta \widetilde {q}_{_S}^{^A}\left({y,t}\right)\ +\
\widetilde {P}_{gg}\left({{x \over y}}\right)\
\delta \widetilde {g}^{^{A}}\left({y,t}\right)\ }\right]
$$
$$
\ \ \ \ \ \ \ \ \ \ \ \ \ \ \ \ \ \ \ \ \ \
\ \ \ \ \ \ \ \ \
+\ \left({3 \over 2}R_A\overline{n}\ -\ {K \over \pi {R_0}^{2}}\right)\
{1 \over {Q}^{2}}\ \left\{{{8{\pi }^{2} \over {N}_{c}^{2}-1} \
\alpha_s (t) \ {C}_{G}^2 \
\int_{x}^{1}{dy \over y}\
{\left[{\widetilde g\left({y,t}\right)}\right]}^{2}}\right\}
\ \ \ .
\eqno{(2.22c)}
$$
where $\bar n$ is the number density $\bar n = 3A/(4\pi R_A^3)$.
Nuclear radius is given by $R_A=R_1A^{1/3}$ with $R_1=1.1$ fm.
For simplicity, we take $R_0$=$R_1$=1 fm and $K$=9/8
so that the nuclear correction terms in Eqs. (2.22a, b, c)
vanish at $A=1$.
We note that two parton number density is
$T_{12}(x_1,x_2,Q^2) = (3/2) R_A \bar n p_1(x_1) p_2(x_2)$.
Therefore, the overall A dependence is $R_A \bar n \propto A^{1/3}$,
which is just the number of nucleons in the longitudinal direction.

\vfill\eject
\vspace{1.0cm}
\noindent
{\Large\bf {3. Brute-force method}}
\vspace{0.4cm}

It is useful to have a computer program of solving the evolution
equations accurately because they are frequently used in theoretical
and experimental studies.
There are a number of methods such as Mellin-transformation and
orthogonal-polynomial methods \cite{SOLVE}.
We have been studying a Laguerre-polynomial method and
it is very efficient by considering computing time and
numerical accuracy \cite{LAG}.
However, the results are slightly worse in the nonsinglet case,
particularly at small and large $x$.
Furthermore, the evolution equations with the recombinations have
complex nonlinear terms, which cannot be handled properly
by the Laguerre polynomials.
In addition to these issues, parton distributions
with singular behavior at small $x$
($xp(x)\rightarrow\infty$ as $x\rightarrow 0$)
could not be evolved either in the program LAG1 nor in LAG2NS
\cite{LAG}.
The singular behavior in sea-quark and gluon distributions
attracts much attention recently due
to the HERA $F_2$ experimental data.

Taking these difficulties into consideration, we decide to employ
a brute-force method \cite{BF}.
The variables $x$ and $t$ are divided into small steps and
integration and differentiation are defined by
$df(x)/dx=[f(x_{m+1})-f(x_m)]/\Delta x_m$ and
$\displaystyle{\int dx f(x)=\sum_{m=1}^{N_x} \Delta x_m}$
$f(x_m)$.
Then, the evolution equations could be solved rather easily.
For example, Eq. (2.3) is written in the following form:
$$
{\widetilde{q}}_{_{NS}}({x}_{k},{t}_{j+1})\
=\ {\widetilde{q}}_{_{NS}}({x}_{k},{t}_{j})\
+\ \Delta {t}_{j}\ \sum\limits_{m=k}^{{N}_{x}}
{\Delta {x}_{m} \over {x}_{m}}
\ {\widetilde{P}}_{_{NS}}\left({{{x}_{k} \over {x}_{m}}}\right)\
{\widetilde{q}}_{_{NS}}({x}_{m},{t}_{j})
\ \ \ .
\eqno{(3.1)}
$$
If the initial distribution ${\widetilde{q}}_{_{NS}}({x}_{m},{t}_{j=1})$
is given, the distribution at $t_1=\Delta t$ is calculated in the above
equation. Repeating this step $N_t-1$ times, we obtain the final distribution
at $t_{N_t}$. This method is very simple but $N_x$ and $N_t$ have to be
large enough to get accurate results.
However, we can reasonably expect that $N_t$ does not have to be
very large. This is because the scaling of
structure functions works approximately, and they depend on the variable
$t$ weakly.
On the other hand, $N_x$ has to be fairly large.
Numerical problems are expected at small $x$
if the $x$ step is taken in the linear scale ($\Delta x=1/N_x$).
The small $x$ region becomes increasingly important with the development
of high-energy accelerators such as HERA.
So it is necessary to have a good numerical method at small $x$ as
small as $10^{-4}$. In order to satisfy this condition,
the logarithmic-$x$ step
$\Delta (log_{10} x)=|log_{10} x_{min}|/N_x$ is
taken in our analysis.

The singlet and $q_i^+$ evolution equations
in Eqs. (2.5a), (2.5b), and (2.6) are more complex,
but they can be solved in the similar way.
These equations are written in the brute-force method as:
\vfill\eject
$$
{\widetilde{q}}_i^{\ +}({x}_{k},{t}_{j+1})\
=\ {\widetilde{q}}_i^{\ +} ({x}_{k},{t}_{j})\
+\ \Delta {t}_{j}\ \sum\limits_{m=k}^{{N}_{x}}
{\Delta {x}_{m} \over {x}_{m}}
\ {\widetilde{P}}_{q^+,\ NS}\left({{{x}_{k} \over {x}_{m}}}\right)\
{\widetilde{q}}_i^+({x}_{m},{t}_{j})
\ \ \ \ \ \ \ \ \ \ \ \ \ \ \ \ \ \ \ \ \ \ \ \ \
$$
$$
+\ \Delta {t}_{j}\ \sum\limits_{m=k}^{{N}_{x}}
{\Delta {x}_{m} \over {x}_{m}}
\ 2C_F T_R {\widetilde{F}}_{qq}\left({{{x}_{k} \over {x}_{m}}}\right)\
{\widetilde{q}}_{_{S}}({x}_{m},{t}_{j})
+\ \Delta {t}_{j}\ \sum\limits_{m=k}^{{N}_{x}}
{\Delta {x}_{m} \over {x}_{m}}
\ {\widetilde{P}}_{qg}\left({{{x}_{k} \over {x}_{m}}}\right)\
{\widetilde{g}}({x}_{m},{t}_{j})
\ \ \ ,
\eqno{(3.2a)}
$$
$$
{\widetilde{q}}_{_{S}}({x}_{k},{t}_{j+1})\
=\ {\widetilde{q}}_{_{S}}({x}_{k},{t}_{j})\
+\ \Delta {t}_{j}\ \sum\limits_{m=k}^{{N}_{x}}
{\Delta {x}_{m} \over {x}_{m}}
\ \left[ \
{\widetilde{P}}_{q^+,\ NS}\left({{{x}_{k} \over {x}_{m}}}\right)\ + \
 2 N_f C_F T_R \widetilde F_{qq} \left({{{x}_{k} \over {x}_{m}}}\right)\
\right] \
{\widetilde{q}}_{_{S}}({x}_{m},{t}_{j})
$$
$$
\ \ \ \ \ \ \ \ \ \ \ \ \ \ \ \ \ \ \ \ \ \ \ \ \ \ \ \ \ \ \ \ \ \
\ \ \ \ \ \ \ \ \ \ \ \ \ \ \
+\ \Delta {t}_{j}\ \sum\limits_{m=k}^{{N}_{x}}
{\Delta {x}_{m} \over {x}_{m}}
\ {\widetilde{P}}_{qg}\left({{{x}_{k} \over {x}_{m}}}\right)\
{\widetilde{g}}({x}_{m},{t}_{j})
\ \ \ ,
\eqno{(3.2b)}
$$
$$
{\widetilde{g}}({x}_{k},{t}_{j+1})\
=\ {\widetilde{g}}({x}_{k},{t}_{j})\
+\ \Delta {t}_{j}\ \sum\limits_{m=k}^{{N}_{x}}
{\Delta {x}_{m} \over {x}_{m}}
\ {\widetilde{P}}_{gq}\left({{{x}_{k} \over {x}_{m}}}\right)\
{\widetilde{q}}_{_{S}}({x}_{m},{t}_{j})
$$
$$
\ \ \ \ \ \ \ \ \ \ \ \ \ \ \ \ \ \ \ \ \ \ \ \ \ \ \ \ \ \ \ \ \ \
\ \ \ \ \ \ \ \ \ \ \ \ \ \ \
+\ \Delta {t}_{j}\ \sum\limits_{m=k}^{{N}_{x}}
{\Delta {x}_{m} \over {x}_{m}}
\ {\widetilde{P}}_{gg}\left({{{x}_{k} \over {x}_{m}}}\right)\
{\widetilde{g}}({x}_{m},{t}_{j})
\ \ \ .
\eqno{(3.2c)}
$$
If the initial distributions,
$\widetilde q_i^+(x_m,t_{j=0})$,
${\widetilde{q}}_{_{S}}({x}_{m},{t}_{j=0})$,
and ${\widetilde{g}}({x}_{m},{t}_{j=0})$ are provided, evolved
distributions at $t_1=\Delta t$ are calculated in the above equations.
Repeating this step $N_t-1$ times, we obtain the final
evolved distributions.
The other evolution equations in section 2
can be solved in the same way.

We did not write explicitly the integrals associated
with $1/(1-x)_+$ terms in Eq. (3.2). However, special care has
to be taken in calculating
the $1/(1-x)_+$ terms in the splitting functions.
The $+$ function is defined in the integral region
($0\le x \le 1$) [Eq. (B.2)], so that the integral is given
in the brute-force method as
$$
\int_x^1 dx' {{f(x')} \over {(1-x')_+}} \ = \
\sum\limits_{m=k}^{N_x} \Delta x_m {{f(x_m)-f(1)} \over {1-x_m}}
\ + \ f(1) \ln (1-x_k)
\ \ \ .
\eqno{(3.3)}
$$
The integral with $[\ln (1-x)/(1-x)]_+$ is given
in the similar way:
$$
\int_x^1 dx' f(x') \left[ {{\ln (1-x')} \over {1-x'}}\right]_+  =
\sum\limits_{m=k}^{N_x} \Delta x_m
[f(x_m)-f(1)] {{\ln (1-x_m)} \over {1-x_m}}
\ + \ {1 \over 2} f(1) \ln ^2 (1-x_k)
\ \ \ .
\eqno{(3.4)}
$$

While the nonlinear recombination terms
are difficult to be handled in the Laguerre method,
it is straightforward to include them in the brute-force method.
In addition, there is no evolution problem due to
the singular behavior of parton distributions at small $x$,
because all calculations in Eqs. (3.1) and (3.2) could be done
in the $x$ region, $x\ge x_{min}$.
$x_{min}$ is set for $10^{-4}$ in our analysis,
but it could be varied depending on one's interest.
Because the recombination equations in section 2.2 are solved in
the similar way, we do not write corresponding expressions
in the brute-force method.

\vfill\eject
\vspace{1.0cm}
\begin{tabbing}
{\Large\bf 4.} $~$ \= {\Large\bf Description of input parameters}       \\
$~~~$             \> {\Large\bf and input distributions}
\end{tabbing}
\vspace{0.4cm}

For running the FORTRAN-77 program BF1, a user should supply
twenty-one input parameters from the file \#10.
In addition, an input distribution(s) should be
given in a function subroutine(s) in the end of the FORTRAN program
or in an input data file(s), \#13, \#14, \#15, \#16, and/or \#17.
Evolution results are written in the output file \#11.
We explain the input parameters and the input distributions
in the following.

\vspace{1.0cm}
\noindent
{\it 4.1 Input parameters}
\vspace{0.4cm}

There are twenty-one input parameters.
Numerical values of the parameters should be supplied
in the file \#10, then these are read
in the main program.

\noindent
\begin{tabbing}
 (99) \= IORDER   \= = 1, \= structure function \= $xF_1(x,Q^2)$; \kill
 (1) \> IOUT      \> = 1, \>
             {write $x$ and $xq_{_{NS}}(x)$ [or $xF_{_{NS}}(x)$]
              at fixed $Q^2$ (=Q2) in the file \#11;} \\
     \>          \> = 2, \>
             {write $Q^2$ and $xq_{_{NS}}(Q^2)$ [or $xF_{_{NS}}(Q^2)$]
              at fixed $x$ (=XX) in the file \#11;} \\
     \>          \> = 3, \>
             {$x$, $xq_{_S}(x)$ [or $xF_{_{S}}(x)$], and $xg(x)$;} \\
     \>          \> = 4, \>
             {$Q^2$, $xq_{_S}(Q^2)$ [or $xF_{_{S}}(Q^2)$], and $xg(Q^2)$;} \\
     \>          \> = 5, \>
             {$x$, $xq_i^+(x)$ [or $xF_i^+(x)$],
                     $xq_{_S}(x)$ [or $xF_{_{S}}(x)$], and $xg(x)$;} \\
     \>          \> = 6, \>
             {$Q^2$, $xq_i^+(Q^2)$ [or $xF_i^+(Q^2)$],
                     $xq_{_S}(Q^2)$ [or $xF_{_{S}}(Q^2)$], and $xg(Q^2)$.} \\
 (2) \> INPUT    \> = 1, \>
                        {parton distributions in the nucleon;} \\
     \>          \> = 2, \>
                        {parton distributions in a nucleus.} \\
     \>          \>     \> Note: If MODIFY=2 and INPUT=2 are chosen,
                           parton distributions \\
     \>          \>     \> both in a nucleus and in the nucleon should
                           be supplied. \\
 (3) \> IREAD   \> = 1, \> give initial distribution(s)
                                  in function subroutine(s); \\
     \>         \> = 2, \> read initial distribution(s) from data file(s). \\
 (4) \> MODIFY  \> = 1, \>
                        {Altarelli--Parisi $Q^2$ evolution;} \\
     \>         \> = 2, \>
                        {$Q^2$ evolution
                         with parton recombinations (Mueller-Qiu).} \\
 (5) \> INDIST  \> = 1, \>
                        {do not write initial distribution(s);} \\
     \>         \> = 2, \>
                        {write initial distribution(s) in the file \#12;} \\
     \>         \> = 3, \>
                        {write initial distribution(s) in the file \#12
                         without calculating evolution.} \\
 (6) \> IORDER  \> = 1, \>
                        {leading order in $\alpha_ s$;} \\
     \>         \> = 2, \>
                        {next-to-leading order.} \\
 (7) \> ITYPE   \> = 1, \> structure function \> $xF_1(x,Q^2)$; \\
     \>         \> = 2, \>                    \> $~~F_2(x,Q^2)$; \\
     \>         \> = 3, \>                    \> $xF_3(x,Q^2)$; \\
     \>         \> = 4, \>
                        {quark distribution $xq(x,Q^2)$.} \\
     \>         \>      \>
                         ITYPE determines the output distribution type. \\
\end{tabbing}

\vfill\eject
\begin{tabbing}
 (99) \= IORDER   \= = 1, \= structure function \= $xF_1(x,Q^2)$; \kill
 (8) \> ISTEP   \> = 1, \>
                         one-step evolution: \=
                             $xq(Q_0^2) \ \rightarrow \ xq(Q^2)$
                         [or $xF(Q_0^2) \ \rightarrow \ xF(Q^2)$], \\
     \>         \>      \>     \> {$xg(Q_0^2) \ \rightarrow \ xg(Q^2)$;} \\
     \>         \>      \>
                     note: one-step $xF_{1,S}$ evolution is not supplied; \\
     \>         \> = 2, \>
                         two-step evolution: \=
                             $xq(Q_0^2) \ \rightarrow \ xq(Q^2)
                                        \ \rightarrow \ xF(Q^2)$, \\
     \>         \>      \>     \> {$xg(Q_0^2) \ \rightarrow \ xg(Q^2)$.} \\
 (9) \> IMORP   \> = 1, \>
                        {$q_i-\bar q_i$ type distribution;} \\
     \>         \> = 2, \>
                        {$q_i+\bar q_i$ type distribution.} \\
 (10) \> Q02     \>
                {= initial $Q^2$ ($\equiv Q_0^2$ in GeV$^2$)
                   at which an initial distribution is supplied.} \\
(11) \> Q2      \>
                {= $Q^2$ to which the distribution is evolved
                                        ($Q^2\ne Q_0^2$).} \\
(12) \> DLAM    \>
                {= QCD scale parameter $\Lambda_{QCD}$ in GeV.} \\
(13) \> NF      \>
                {= number of quark flavors (NF=3 or 4).} \\
(14) \> XX      \>
                {= $x$ at which $Q^{2}$ dependent
                   distributions are written (IOUT=2, 4, or 6 case).} \\
(15) \> NX      \>
                {= number of $x$ steps (NX$<$5000).} \\
(16) \> NT      \>
                {= number of $t$ steps (NT$<$5000).} \\
(17) \> NSTEP  \>
                {= number of $x$ steps or $t$ steps
                           for writing output distribution(s).} \\
(18) \> NXMIN   \>
                {= $log_{10}$(minimum of $x$)
                   $[0<min(x)= 10^{^{NXMIN}}<XX]$.} \\
(19) \> NA      \>
                {= mass number of a nucleus (NA=1 in the nucleon case).} \\
(20) \> RFM     \>
                {= $R_0$=$R_1$ (fm) in Eqs. (2.19) and (2.22).} \\
(21) \> DKHT    \>
                {= $K_{HT}$ in $G_{HT}(x)$ in Eq. (2.20).} \\
     \>         \>
            {\ \ ($G_{HT}(x)$ is a higher dimensional gluon distribution).} \\
\end{tabbing}

\vspace{-0.4cm}
The meaning of IREAD is explained in section 4.2.
IMORP=1 or 2 means that the distribution is
$\displaystyle{\sum_i a_i (q_i+\bar q_i)}$ or
$\displaystyle{\sum_i a_i (q_i-\bar q_i)}$ type,
where $a_i$ are constants.
For example, $u_v+d_v$=$(u-\bar u)+(d-\bar d)$ and
$F_3^{\nu p}+F_3^{\bar\nu p}$ are $q_i-\bar q_i$ type distributions.
$F_2^{ep}$ is obtained by the convolution of the distributions,
$(4/9)x(u+\bar u+c+\bar c)$$+(1/9)x(d+\bar d+s+\bar s)$ and $xg$,
with the corresponding coefficient functions, so that it is
a $q_i+\bar q_i$ type distribution.
In the evolution of nucleon structure functions, NA=1 should be supplied.
In the Altarelli-Parisi evolution, the constants $RKM$ and
$DKHT$ are unnecessary so that arbitrary constants may be supplied.

For example, if one would like to evolve an initial
singlet-quark distribution $xq_{_S}$ at $Q^2$=4 GeV$^2$
in the nucleon to the singlet structure function
$F_{2,S}$ at $Q^2$=200 GeV$^2$
by the NLO Altarelli-Parisi with $N_f$=4 and $\Lambda$=0.255 GeV,
the input parameters could be
IOUT=3, INPUT=1, IREAD=1, MODIFY=1, INDIST=1, IORDER=2,
ITYPE=2, ISTEP=2, IMORP=2, Q02=4.0, Q2=200.0, DLAM=0.255,
NF=4, XX=0.0, NX=1000, NT=200, NSTEP=100, NXMIN=$-$4,
NA=1, RFM=0.0, and DKHT=0.0.
In this case,  the input file \#10 is the following.

\vspace{0.3cm}
\noindent
3, 1, 1, 1, 1, 2, 2, 2, 2 \\
4.0, 200.0, 0.255, 4, 0.0, 1000, 200, 100, $-$4, 1, 0.0, 0.0 \\

\vfill\eject
\noindent
{\it 4.2 Input distributions supplied by function subroutines (IREAD=1)}
\vspace{0.4cm}

If IREAD=1 is chosen, an input distribution(s) at $Q_0^2$
should be supplied in the end of the FORTRAN program BF1
as a function subroutine(s).

\vspace{0.3cm}
\noindent
1) Nonsinglet case in the nucleon and nuclei

An initial  nonsinglet-quark distribution
(or a nonsinglet structure function)
at $Q_0^2$ should be given in QNS0(X)
as a double precision function.
As an example, the MRS(G) valence quark distribution
$xu_v+xd_v$ \cite{MRSG} at $Q_0^2$=4 GeV$^2$ is given
in the original program BF1.

\vspace{0.3cm}
\noindent
2) Singlet case in the nucleon

An initial singlet-quark distribution (or a singlet structure function)
in the nucleon
at $Q_0^2$ should be supplied in QS0(X),
and an initial gluon distribution
in the nucleon should be in G0(X).
These subroutines are used in the nucleon case and also
in the nuclear case with MODIFY=2.
The MRS(G) $xq_{_S}=xu_v+xd_v+xS$ and $xg$ distributions are given.

\vspace{0.3cm}
\noindent
3) $q_i^+$ ($F_i^+$) distribution case in the nucleon

An initial $q_i^+$ distribution (or $F_i^+$)
in the nucleon at $Q_0^2$ should be
supplied in QI0(X),
a singlet-quark distribution (or a singlet structure function)
in the nucleon should be in QS0(X),
and an initial gluon distribution in the nucleon should be in G0(X).
These subroutines are used in the nucleon case and also
in the nuclear case with MODIFY=2.
The MRS(G) distributions are given in each function subroutine.
$xq_i^+=xd+x\bar d$ is given as an example.

\vspace{0.3cm}
\noindent
4) Singlet case in a nucleus

An initial nuclear singlet-quark distribution
(or a singlet structure function)
at $Q_0^2$ should be supplied in QSA0(X),
and a nuclear gluon distribution should be in GA0(X).
These subroutines are used only in the nuclear case.
The SK $xq_{_S}^{{Ca}}$ and $xg^{{Ca}}$ distributions \cite{SKF2}
are given as an example.

\vspace{0.3cm}
\noindent
5) $q_i^+$ ($F_i^+$) distribution case in a nucleus

An initial nuclear $q_i^+$ distribution (or $F_i^+$)
at $Q_0^2$ should be supplied in QIA0(X),
a nuclear singlet-quark distribution (or a singlet structure function)
should be in QSA0(X),
and an nuclear gluon distribution should be in GA0(X).
These subroutines are used only in the nuclear case.
The SK distributions are given in each function subroutine.

\vfill\eject
\noindent
{\it 4.3 Input distributions supplied by data files (IREAD=2)}
\vspace{0.4cm}

If IREAD=2 is chosen, an input distribution(s) at $Q_0^2$
should be supplied in a separate data file(s).

\vspace{0.3cm}
\noindent
1) Nonsinglet case in the nucleon and nuclei (data file \#13)

An initial  nonsinglet-quark distribution
(or a nonsinglet structure function)
at $Q_0^2$ should be given in the data file \#13
as shown in the following example.

\begin{tabbing}
      0.000100 \ \ \ \ \ \ \= 0.023868 \kill
      0.000100   \>  0.023868 \\
      0.000110   \>  0.024942 \\
      0.000120   \>  0.026069 \\
      0.000132   \>  0.027251 \\
      0.000145   \>  0.028491 \\
       ...       \>   ...     \\
       ...       \>   ...     \\
       ...       \>   ...     \\
      1.000000   \>  0.000000 \\
\end{tabbing}

\vspace{-0.3cm}
The first column is the $x$ values and the second
one is the corresponding $xu_v+xd_v$ values.
The data at $x\le x_{min}$ and at $x$=1.0 must be supplied.

\vspace{0.3cm}
\noindent
2) Singlet case in the nucleon (data file \#14)

An initial singlet-quark distribution (or a singlet structure function)
and a gluon distribution should be given in the data file \#14
as shown in the following.

\begin{tabbing}
      0.000100 \ \ \ \ \ \ \= 3.137921 \ \ \ \ \ \ \= 23.163731 \kill
      0.000100   \> 3.137921  \> 23.163731 \\
      0.000110   \> 3.114648  \> 22.485543 \\
      0.000120   \> 3.091368  \> 21.825121 \\
      0.000132   \> 3.068077  \> 21.181969 \\
      0.000145   \> 3.044768  \> 20.555604 \\
       ...       \>  ...      \>   ...     \\
       ...       \>  ...      \>   ...     \\
       ...       \>  ...      \>   ...     \\
      1.000000   \> 0.000000  \> 0.000000 \\
\end{tabbing}

\vspace{-0.3cm}
The first column is the $x$ values, the second is
the $xq_{_S}$ distribution, and the third is
the $xg$ distribution.
The data at $x\le x_{min}$ and at $x$=1.0 must be supplied.

\vspace{0.3cm}
\noindent
3) $q_i^+$ ($F_i^+$) distribution case in the nucleon (data file \#15)

An initial $q_i^+$ distribution (or $F_i^+$),
a singlet-quark distribution (or a singlet structure function),
and a gluon distribution should be given in the data file \#15
as shown in the following.

\begin{tabbing}
0.000100 \ \ \ \ \ \ \= 1.229377 \ \ \ \ \ \ \= 3.137921
\ \ \ \ \ \ \= 23.163731 \kill
      0.000100   \> 1.229377   \> 3.137921  \> 23.163731 \\
      0.000110   \> 1.220390   \> 3.114648  \> 22.485543 \\
      0.000120   \> 1.211413   \> 3.091368  \> 21.825121 \\
      0.000132   \> 1.202445   \> 3.068077  \> 21.181969 \\
      0.000145   \> 1.193483   \> 3.044768  \> 20.555604 \\
       ...       \>  ...       \>  ...      \>   ...     \\
       ...       \>  ...       \>  ...      \>   ...     \\
       ...       \>  ...       \>  ...      \>   ...     \\
      1.000000   \> 0.000000   \> 0.000000  \>  0.000000 \\
\end{tabbing}

\vspace{-0.3cm}
The first column is the $x$ values,
the second is the $xu^+$ distribution,
the third is $xq_{_S}$,
and the fourth is $xg$.
The data at $x\le x_{min}$ and at $x$=1.0 must be supplied.

\vspace{0.3cm}
\noindent
4) Singlet case in a nucleus (data file \#16)

An initial nuclear singlet-quark distribution
(or a singlet structure function)
and a nuclear gluon distribution should
be given in the data file \#16
as shown in the nucleon case 2).
In the MODIFY=2 case, the nucleon data file \#14 should be
also supplied.

\vspace{0.3cm}
\noindent
5) $q_i^+$ ($F_i^+$) distribution case in a nucleus (data file \#17)

An initial nuclear $q_i^+$ distribution (or $F_i^+$),
a nuclear singlet-quark distribution (or a singlet structure function),
and a nuclear gluon distribution should be given in the data file \#17
as shown in the nucleon case 3).
In the MODIFY=2 case, the nucleon data file \#15 should be
also supplied.

\vfill\eject
\vspace{1.0cm}
\noindent
{\Large\bf {5. Description of the program BF1}}
\vspace{0.4cm}

\noindent
{\it 5.1 Main program BF1}
\vspace{0.3cm}

The main program reads twenty-one input parameters
from the input file \#10.
The parameters are checked by the subroutine ERR.
If there is an error, the program stops.
Then, $t$ defined in Eq. (2.1) is evaluated.
Spline coefficients of the Spence function in Eqs. (B.7) and (B.8)
are calculated by calling the SPLINE subroutine,
and the interpolated function is used
throughout the program for saving computation time.
In the end, GETQNS is called in the nonsinglet case
and GETQS is called in the singlet (or $xq^+$, $xF^+$) case
for calculating the $Q^2$ evolution.

\vspace{0.4cm}
\noindent
{\it 5.2 Subroutine GETQNS}
\vspace{0.3cm}

First, the initial distribution is taken from
the subroutine QNS0 or is read from the file \#13,
and it is stored in the array QNS(I).
The distribution is written in the file \#12 if INDIST$\neq$1.
The initial distribution is stored in the array WQNS(I).
The evolved distribution at $t_1=\Delta t$ is calculated
by calling QNSXT and it is stored in QNS(I).
The major part of the NS evolution calculation is done
in the QNSXT subroutine.
Next, QNS(I) at $t_1=\Delta t$ is stored in WQNS(I).
The distribution at $t_2=2\Delta t$ is calculated
again by calling QNSXT and it is stored in QNS(I).
Repeating this step, we obtain the final distribution
at $t=N_t \Delta t$.
The results are interpolated either in the variable $t$
or in $x$ and they are written in the output file \#11.

\vspace{0.4cm}
\noindent
{\it 5.3 Subroutine GETQS}
\vspace{0.3cm}

The initial distributions, $xq_{_S}$ and $xg$
($xq_i^+$, $xq_{_S}^{^A}$, $xg^{^A}$, $xq_i^{+,A}$), are taken from
the functions, QS0 and G0 (QI0, QSA0, GA0, QIA0)
or from the file \#14, \#15, \#16, or \#17.
The distributions are stored in the arrays,
QS(I) and G(I) (QI(I), QSA(I), GA(I), QIA(I)), and
they are written in the file \#12 if INDIST$\neq$1.
The initial distributions are stored in WQS(I) and WG(I)
(WQI(I), WDQS(I), WDG(I), WDQI(I), WGHT(I))
[see Eqs. (2.20) and (2.21) for $xG_{HT}$ and $\delta (xp^{^A})$)].
The evolved ones at $t_1=\Delta t$ are calculated by calling GETQGX,
and results are stored in
QS(I) and G(I) (QI(I), QSA(I), GA(I), QIA(I)).
Next, the evolution results are stored in  WQS(I) and WG(I)
(WQI(I), WDSQ(I), WDG(I), WDQI(I), WGHT(I)) and the evolution
from $t_1=\Delta t$ to $t_2=2\Delta t$ is calculated.
Repeating this step, we obtain the final distributions
at $t=N_t \Delta t$.
The results are interpolated either in the variable $t$
or in $x$ and they are written in the output file \#11.

\vspace{0.4cm}
\noindent
{\it 5.4 Subroutines ERR(IERR), INVERT(NN,QQQ)}
\vspace{0.3cm}

The subroutine ERR checks whether the input parameters
are valid.
If they are not valid, this subroutine
returns IERR=1 in the main program.
The subroutine INVERT reverses the array QQQ, namely
QQQ(1)$\rightarrow$QQQ(NN),
QQQ(2)$\rightarrow$QQQ(NN$-$1), ...,
QQQ(NN)$\rightarrow$QQQ(1).
This is used in the $Q^2$ devolution case ($Q^2<Q_0^2$).

\vspace{0.4cm}
\noindent
{\it 5.5 Functions FUNQ2J(IORDER,ALPHA) and ALPHAN(QQ)}
\vspace{0.3cm}

The function FUNQ2J obtains the $Q^2$ from a given running
coupling constants $\alpha_s$ within the range
 from $Q_0^2-0.01$(GeV$^2$) to $Q^2+0.01$(GeV$^2$) if $Q_0^2<Q^2$
(from $Q^2-0.01$ to $Q_0^2+0.01$ if $Q^2<Q_0^2$).
This function is used in calculating $Q^2$ from given $t$.
The function ALPHAN(QQ) obtains the NLO running
coupling constant $\alpha_s$ at $Q^2$ in the $\overline{MS}$ scheme.

\vspace{0.4cm}
\noindent
{\it 5.6 Subroutine GETQGX(ALPHA)}
\vspace{0.3cm}

The subroutine GETQGX calculates the singlet (and $q_i^+$, $F_i^+$)
evolution by calling functions, QSXT, GXT, QSKXT, GKXT, GHTXT,
DQSXT, and DGXT, which corresponds to the evolution of
$xq_{_S}$ ($xq_i^+$), $xg$, $xq_{_S}$(recomb.), $xg$(recomb.),
$xG_{HT}$, $\delta (xq_{_S})$ ($\delta (xq_i^+)$), and
$\delta (xg)$.
(or structure functions) respectively.
In the Altarelli-Parisi evolution in the nucleon, only
QSXT and GXT are called. In addition to these,
QSKXT, GKXT, and GHTXT are called in the Mueller-Qiu case.
DQSXT and DGXT are called in the nuclear case.

\vspace{0.4cm}
\noindent
{\it 5.7 Subroutine DELTA0(QGA,QG,DQG)}
\vspace{0.3cm}

This subroutine calculates the difference DQG between
the nucleon parton distribution QG and the nuclear one QGA.

\vspace{0.4cm}
\noindent
{\it 5.8 Functions QSKXT(I,FF) and GKXT(I)}
\vspace{0.3cm}

These functions calculate recombination effects in the nucleon.
QSKXT calculates the recombination terms in
$xq_{_S}$ or $xq_i^+$ (or structure functions),
and GKXT calculates the recombination in $xg$.
The flavor number FF is used for obtaining
either $xq_{_S}$ (FF=$N_f$) or $xq_i^+$ (FF=1).

\vspace{0.4cm}
\noindent
{\it 5.9 Functions QNSXT(I,ALPHA,IORDER,ITYPE,SIGN),}

\ \ \ \ \ \ \ \ \ \ \ \ \ \  {\it QSXT(I,ALPHA,QSK,WWQ,IORS,FF),
                                  GXT(I,ALPHA,GK),}

\ \ \ \ \ \ \ \ \ \ \ \ \ \  {\it DQSXT(I,ALPHA,WWQ,IORS,FF),
                                  DGXT(I,ALPHA)}
\vspace{0.3cm}

These function calculates the $Q^2$ evolution
from $t_j$ to $t_j+\Delta t$.
QNSXT, QSXT, GXT, DQSXT, and DGXT calculate
the evolution of nonsinglet-quark, singlet-quark,
gluon, nuclear (actually, nuclear correction of) nonsinglet-quark,
nuclear gluon distributions
(and corresponding structure functions) respectively.
The evolution in the brute-force method
is discussed in detail in section 3.
If IORS=1, the singlet distribution is calculated, and
the $xq_i^+$ ($xF_i^+$) distribution is calculated if IORS=2.

\vspace{0.4cm}
\noindent
{\it 5.10 Functions PNS0(I,K,ZK) and PNS1(I,K,ZK,SIGN,WWQ)}
\vspace{0.3cm}

The function PNS0 (PNS1) calculates the LO (NLO) nonsinglet
splitting function multiplied by a nonsinglet quark distribution
or a structure function, which is stored in WWQ.
If IMORP=1 (SIGN=$-$1.0), PNS1 uses the ``$q-\bar q$" type splitting
function, and it uses the ``$q+\bar q$" type splitting function
if IMORP=2 (SIGN=+1.0).

\vspace{0.4cm}
\noindent
{\it 5.11 Functions PSQ0(I,K,ZK,WWQ,WWG,FF), PSG0(I,K,ZK,WWQ,WWG),}

\ \ \ \ \ \ \ \ \ {\it PSQ1(I,K,ZK,WWQI,WWQ,WWG,FF), and
                                 PSG1(I,K,ZK,WWQ,WWG)}
\vspace{0.3cm}

These functions calculate splitting functions multiplied by
quark and gluon distributions.
The function PSQ0 calculates
$\tilde{P}_{qq}^{(0)}(\frac{x}{y})WWQ
+\tilde{P}_{qg}^{(0)}(\frac{x}{y})WWG$,
where WWQ is $yq_{_{S}}(y)$, $yF_{_{S}}(y)$, $yq_i^+(y)$,
or $yF_i^+(y)$
and WWG is $yg(y)$.
The function PSG0 calculate
$\tilde{P}_{gq}^{(0)}(\frac{x}{y})WWQ
+\tilde{P}_{gg}^{(0)}(\frac{x}{y})WWG$.
PSQ1 and PSG1 calculates the same quantities in NLO.

\vspace{0.4cm}
\noindent
{\it 5.12 Functions CQ1(I,K,ZK,ITYPE,WWQ) and CG1(K,ZK,ITYPE,WWG)}
\vspace{0.3cm}

The function CQ1 calculates the NLO coefficient function
${C}_{q}^{(1)}$ multiplied by
a quark distribution or a structure function (WWQ).
CG1 calculates ${C}_{g}^{(1)}$ multiplied by
a gluon distribution (WWG).
${C}_{q}^{(1)}$ and ${C}_{g}^{(1)}$ are
denoted $B_n^q$ and $B_n^g$ in Appendix C.

\vspace{0.4cm}
\noindent
{\it 5.13 Functions EQQ(K,ZK,WWQ), EQG(K,ZK,WWG) and EGQ(K,ZK,WWQ)}
\vspace{0.3cm}

These functions calculate $E_{qq}$, $E_{qg}$ and $E_{gq}$ multiplied
by a structure function or a gluon distribution
in the one-step evolution.
$E_{gg} = - E_{qq}$ is used in calculating the $E_{gg}$ part.
See Appendix B for the explicit expressions of
$E_{qq}$, $E_{qg}$ and $E_{gq}$.

\vspace{0.4cm}
\noindent
{\it 5.14 Function GHTXT(I)}
\vspace{0.3cm}

This function calculates the evolution of
the higher dimensional gluon distribution $xG_{_{HT}}$
from $t_j$ to $t_j+\Delta t$ in Eq. (2.19d).

\vspace{0.4cm}
\noindent
{\it 5.15 Subroutines GETFNS(ITYPE,ALPHA,QNS,FNS),}

\ \ \ \ \ \ \ \ \ \ \ \ \ \ \ \ \ \ {\it GETF(IORS,ITYPE,ALPHA,QS,G,FS)}

\vspace{0.3cm}

These subroutines calculate structure functions
from quark and gluon distributions
by taking convolutions with the coefficient functions.
These are used in the two-step evolution.
GETFNS obtains a nonsinglet structure function FNS
from a nonsinglet quark distribution QNS.
GETF obtains a singlet structure function (or $xF_i^+$) FS
from singlet quark (or $xq_i^+$) and gluon distributions, QS and G.

\vfill\eject
\noindent
{\it 5.16 Function SPENCE(X)}
\vspace{0.3cm}

This function calculates the Spence function
by using a series expansion form in Eq. (B.8).
The upper limit of the summation is taken as max(k)=10000
so that accuracy of the obtained function is better than $10^{-5}$.
The Spence function appears in the Splitting functions
and it could take significant computing time if the function
SPENCE is repeated called. Therefore, the interpolated one is used
in calculating the splitting functions.

\vspace{0.4cm}
\noindent
{\it 5.17 Subroutines DATAR1(N,QFX), DATAR2(N,QFX,GGX),}

\ \ \ \ \ \ \ \ \ \ \ \ \ \ \ \ \ \ {\it DATAR3(N,QIX,QFX,GGX)}

\vspace{0.3cm}

IF IREAD=2 is chosen, an initial distribution is read
from a data file. DATAR1 reads an initial nonsinglet
distribution form the data file \#N.
DATAR2 reads an initial singlet-quark distribution $xq_{_S}$
(or structure function $xF_{_S}$) and a gluon distribution $xg$.
DATAR3 reads $xq_i^+$ ($F_i^+$), $xq_{_S}$ ($xF_{_S}$), and $xg$.
Then, the initial distribution data are interpolated, so that
the reasonably large amount of data should be supplied in the file.

\vspace{0.4cm}
\noindent
{\it 5.18 Subroutine SPLINE(N,X,Y,B,C,D)
          and function SEVAL(N,XX,X,Y,B,C,D)}
\vspace{0.3cm}

The subroutine SPLINE calculates the coefficients, B(I), C(I), and D(I)
($I=1,2, \cdot\cdot\cdot, N$) in a cubic Spline interpolation.
Y(I) at X(I) ($I=1,2, \cdot\cdot\cdot, N$) are supplied.
This interpolation program is taken from Ref. \cite{FMM}.
Using the obtained Spline coefficients, the function SEVAL calculates
the value of Y at given XX.

\vspace{0.4cm}
\noindent
{\it 5.19 Functions QQ(X,A,B,C), QQMRS(X,A,B,C,D,E)}
\vspace{0.3cm}

These functions calculate typical parton distributions given
by the parameters A, B, C, D, and E.
QQ calculates $Ax^B(1-x)^C$, and
QQMRS does $Ax^B(1+C\sqrt{x}+Dx)(1-x)^E$.

\vspace{0.4cm}
\noindent
{\it 5.20 Functions QNS0(X), QS0(X), and G0(X)}
\vspace{0.3cm}

The functions QNS0, QS0, and G0 calculate
an initial nonsinglet-quark distribution $xq_{_{NS}}$ ($xF_{_{NS}}$),
a singlet-quark distribution $xq_{_S}$ ($xF_{_S}$), and a gluon
distribution $xg$ in the nucleon.
As an example, the MRS(G) distributions \cite{MRSG} are provided.
The nonsinglet one is
QNS0=$xu_v+xd_v$=$2.704x^{0.593}$$(1-0.76\sqrt{x}
+4.20x)$$(1-x)^{3.96}$+$0.2513x^{0.335}$$(1+8.63\sqrt{x}
+0.32x)$$(1-x)^{4.41}$,
the singlet one is
QS0=$xu_v+xd_v+xS$=$2.704x^{0.593}$$(1-0.76\sqrt{x}
+4.20x)$$(1-x)^{3.96}$+$0.2513x^{0.335}$$(1+8.63\sqrt{x}
+0.32x)$$(1-x)^{4.41}$+$1.74 x^{-0.067}$$(1-3.45\sqrt{x}
+10.3x)$$(1-x)^{10.1}$, and the gluon distribution is
G0=$1.51x^{-0.301}$$(1-4.14\sqrt{x}+10.1x)$$(1-x)^{6.06}$.

\vfill\eject
\noindent
{\it 5.21 Functions QSA0(X) and GA0(X)}
\vspace{0.3cm}

The functions QSA0 and GA0 calculate
a singlet-quark distribution $xq_{_S}$ ($xF_{_S}$) and a gluon
distribution $xg$ in a nucleus.
As an example, the SK distributions \cite{SKF2}
at $Q_0^2$=0.8 GeV$^2$ are provided.
The singlet one is
QSA0=$1.840x^{0.472}$$(1-0.984x)^{4.06}$$(1
+9.33x)$+$6.423x^{0.600}$$(1-x)^{8.13}$$(1-0.568x)$, and
the gluon distribution is
GA0=$179.2x^{1.95}$$(1-x)^{7.32}$$(1-0.619x)$.

\vspace{0.4cm}
\noindent
{\it 5.22 Functions QI0(X) and QIA0(X)}
\vspace{0.3cm}

The function QI0 or QIA0 calculates
an initial distribution $xq_i^+$ ($xF_i^+$)
in the nucleon or the one in a nucleus.
As an example, the MRS(G) and SK $xd^+$ distributions are provided.
The nucleon one is
QI0=+$0.2513x^{0.335}$$(1+8.63\sqrt{x}
+0.32x)$$(1-x)^{4.41}$+$0.4(1-0.02)$$1.74 x^{-0.067}$$(1-3.45\sqrt{x}
+10.3x)$$(1-x)^{10.1}$$+$$0.043x^{0.3}(1-x)^{10.1}(1+64.9x)$, and
the nuclear one is
QIA0=$1.773x^{0.656}$$(1-0.956x)^{5.23}$$(1
+3.01x)$+$6.423x^{0.600}$$(1-x)^{8.13}$$(1-0.568x)/3$.

\vfill\eject
\vspace{1.0cm}
\noindent
{\Large\bf {6. Numerical analysis}}
\vspace{0.4cm}

Parton distributions or structure functions either
in the nucleon or in a nucleus could be evolved
in the program BF1.
We discuss numerical results on both cases.

\vspace{0.6cm}
\noindent
{\it 6.1 Accuracy of $Q^2$ evolution results}
\vspace{0.4cm}

We check the LO and NLO splitting functions,
the coefficient functions, and the function ${\bf E}$
by comparing their moments with other calculations
in Refs. \cite{HWW,FRS,BBDM}.
Because the different convention is employed for the gluon
distribution from the one in Ref. \cite{HWW},
${\bf D}^{(1)}$ and ${\bf E}$ in Eqs. (2.17) and (2.18)
are slightly different.
We compare BF1 evolution results with the evolution of HMRS-B
\cite{HMRS}. Choosing the HMRS-B singlet-quark and gluon distributions
at $Q_0^2$=4 GeV$^2$ as the initial distributions, we calculate
the distributions at $Q^2$=200 GeV$^2$ and compare them with
the HMRS-B evolution results.
We find that both agree well in the sense
that the differences are typically within 2\% in the region ($0.01<x$).
However, the differences become slightly larger at small $x$
(2$-$4\% at $0.0001<x<0.001$).
We also checked various BF1 evolution results
by comparing with those in the Laguerre polynomial method \cite{LAG}.
It indicates that both results agree within about 1\% accuracy.
Furthermore, we checked that results in the one-step evolution
[$xF(x,Q_0^2)$$\rightarrow$$xF(x,Q^2)$]
agree well with those in the two-step evolution
[$xq(x,Q_0^2)$$\rightarrow$$xq(x,Q^2)$$\rightarrow$$xF(x,Q^2)$].
{}From these comparisons and repeated checks on the program,
the program BF1 is believed to be a reliable evolution program.
We discuss the details of numerical accuracy
in the following.

First, $Q^2$ evolution of a nonsinglet distribution
is calculated in the program BF1.
There are essentially two parameters which determine
numerical accuracy of the $Q^2$ evolution.
These are numbers of points in variables $x$ and $t$ ($N_x$, $N_t$).
We show dependency of our results on these numbers.
For running the program BF1, the input parameters and
an input distribution in section 4 should be supplied.
As an initial  nonsinglet distribution, we employ the HMRS-B
$xu_v+xu_d$ at $Q_0^2$=4 GeV$^2$. It is evolved up to
$Q^2$=200 GeV$^2$ with $\Lambda$=0.19 GeV and $N_f$=4.

In Fig. 1a, $N_x$ is fixed at $N_x$=1000, then $N_t$=10, 20, or 500
is taken for finding the dependency on $N_t$.
For example, the input parameters in the case of $N_t$=500
are
IOUT=1, INPUT=1, IREAD=1, MODIFY=1, INDIST=2, IORDER=2,
ITYPE=4, ISTEP=1, IMORP=1, Q02=4.0, Q2=200.0, DLAM=0.255,
NF=4, XX=0.0, NX=1000, NT=500, NSTEP=100, NXMIN=$-$4,
NA=1, RFM=0.0, and DKHT=0.0.
There are solid, dashed, and dotted curves at $Q^2$=200 GeV$^2$;
however, these cannot be distinguished in Fig. 1.
It indicates that numerical results are almost independent
of the number $N_t$ in the nonsinglet case if $N_t$ is more
than a certain number. This result is, in fact, anticipated
because the scaling hypothesis works approximately
in parton distributions.
In comparison with the $N_t$=500 results, numerical differences are
within   1\% at $10^{-4}<x<0.5$ and   4\% at $x$=0.8
in the $N_t$=10 case,
within 0.6\% at $10^{-4}<x<0.5$ and   2\% at $x$=0.8 in $N_t$=20, and
within 0.2\% at $10^{-4}<x<0.5$ and 0.8\% at $x$=0.8 in $N_t$=50.

Next, $N_x$ is varied with the fixed $N_t$=50 in Fig. 1b.
Four types of curves are shown at $Q^2$=200 GeV$^2$.
The dotted curve is the evolution result for $N_x$=100,
the dashed for $N_x$=300, the dot-dashed for $N_t$=1000,
and the solid for $N_t$=4000.
As it is shown in the figure, we should take rather larger
number of $N_x$ for getting converging results.
The accuracy is worse if $N_x <$100.
In comparison with the $N_x$=4000 results, numerical differences are
within   9\% at $10^{-4}<x<0.5$ and 24\% at $x$=0.8 in the $N_x$=100 case,
within   3\% at $10^{-4}<x<0.5$ and 8\% at $x$=0.8 in $N_x$=300, and
within 0.7\% at $10^{-4}<x<0.5$ and 2\% at $x$=0.8 in $N_x$=1000.

{}From the above results, we recommend to use
$N_t\ge$50 and $N_x\ge$1000 for getting accuracy better than 1\%
at $0.0001<x<0.5$ and better than 2\% at $0.5<x<0.8$
in the nonsinglet case.
Running CPU time with $N_t$=50 and $N_x$=1000 is about
five minutes on the AlphaServer 2100 4/200.

Singlet distribution results are shown in Fig. 2a, where
$N_t$ is varied with fixed $N_x$=1000.
The initial distributions are the HMRS-B singlet-quark
and gluon distributions at $Q_0^2$=4 GeV$^2$.
The dotted, dashed, dot-dashed, and solid curves are
the results for $N_t$=20, 50, 200, and 1000 respectively.
As it is obvious from the figure, we need to take
large $N_t$ for getting convergent results at small $x$.
In comparison with the $N_t$=1000 results, numerical differences are
within   5\% at $10^{-4}<x<0.9$ in the $N_t$=20 case,
within   2\% at $10^{-4}<x<0.9$ in $N_t$=50, and
within 0.4\% at $10^{-4}<x<0.9$ in $N_t$=200.
In Fig. 2b, $N_x$ is varied with fixed $N_t$=200 in
the singlet distribution. The dotted, dashed, dot-dashed, and solid
curves are obtained by taking $N_x$=100, 300, 1000, and 4000
respectively.
In comparison with the $N_x$=4000 results, numerical differences are
within   8\% at $10^{-4}<x<0.5$ and 24\% at $x$=0.8 in the case $N_x$=100,
within   3\% at $10^{-4}<x<0.5$ and  8\% at $x$=0.8 in $N_x$=300, and
within 0.7\% at $10^{-4}<x<0.5$ and  2\% at $x$=0.8 in $N_x$=1000.

Evolution results of a gluon distribution are shown in Figs. 3a and 3b,
and they show similar tendency to those in Figs. 2a and 2b.
In. Fig 3a, $N_t$ is varied with fixed $N_x$=1000.
In comparison with the $N_t$=1000 results in Fig. 3a,
numerical differences are
within   5\% at $10^{-4}<x<0.6$ in the $N_t$=20 case,
within   2\% at $10^{-4}<x<0.6$ in $N_t$=50, and
within 0.4\% at $10^{-4}<x<0.6$ in $N_t$=200.
In Fig. 3b, $N_x$ is varied with fixed $N_t$=200.
In comparison with the $N_x$=4000 results in Fig. 3b,
numerical differences are
within  25\% at $10^{-4}<x<0.6$ in the case $N_x$=100,
within   8\% at $10^{-4}<x<0.6$ in $N_x$=300, and
within   1\% at $10^{-4}<x<0.2$ and
within   2\% at $0.2<x<0.6$ in $N_x$=1000.
The accuracy is slightly worse in the gluon distribution but
it is mainly in the medium $x$ region, which is not important
in the gluon distribution.

{}From these results, we recommend to use
$N_t\ge$200 and $N_x\ge$1000 in the singlet case
for getting accuracy better than 1\% at $0.0001<x<0.5$
and 2\% at $0.5<x<0.8$ in the singlet-quark distribution,
and accuracy is better than 1\% at $0.0001<x<0.2$ and
2\% at $0.2<x<0.6$ in the gluon distribution.
Typical running time with $N_t$=200 and $N_x$=1000
on the AlphaServer is about one hour
in the NLO singlet case.

We also check numerical accuracy of our evolution results
in the case of $q_i+\bar q_i$ distributions,
in the case with the recombinations, and
in the case of nuclear parton distributions.
However, the results have similar dependency on
$N_t$ and $N_x$, so that they are not discussed in
this paper. A user can always check one's evolution results
by changing $N_t$ and $N_x$ if one is suspicious about
the accuracy.

The one-step evolution,
e.g. $xq_{_S}(x,Q_0^2)$$\rightarrow$$xq_{_S}(x,Q^2)$,
is used for getting the above results.
We check the accuracy in the two-step evolution case,
e.g. $xq_{_S}(x,Q_0^2)$$\rightarrow$$xq_{_S}(x,Q^2)$
$\rightarrow$$xF_{_S}(x,Q^2)$,
by calculating the $F_2$ structure function
of the proton with the MRS(G) input distributions.
We find that the accracy is better than 1\%
in the region ($10^{-4}<x<0.8$) by taking
$N_t$=200 and $N_x$=1000.
Similar results are obtained in the recombination case,
so that we do not discuss their numerical details.

{}From the above discussions,
we had better use $N_t\ge$200 and $N_x\ge$1000
for getting accuracy better than 2\%.
Typical running time could be an hour on the AlphaServer.
If the program is run on a SUN workstation, it could take
a significant amount of running time.
Therefore, a user may use a reasonably powerful machine
in running the program BF1.

\vspace{0.6cm}
\noindent
{\it 6.2 Comparisons with experimental data}
\vspace{0.4cm}

We first compare nonsinglet evolution results with
CDHSW $F_3$-structure-function data \cite{CDHSW} in neutrino reactions.
We choose the initial distribution as the MRS(G) $xu_v+xd_v$
distribution at $Q_0^2$=4 GeV$^2$ \cite{MRSG},
$xu_v+xd_v$=$2.704x^{0.593}$$(1-0.76\sqrt{x}
+4.20x)$$(1-x)^{3.96}$+$0.2513x^{0.335}$$(1+8.63\sqrt{x}
+0.32x)$$(1-x)^{4.41}$,
The two-step part of the evolution program is used
for calculating $xF_3(x,Q^2)$ from the initial
$xu_v(x,Q_0^2)$$+xd_v(x,Q_0^2)$.
Chosen input parameters are the following in the NLO case:
IOUT=2, INPUT=1, IREAD=1, MODIFY=1, INDIST=1, IORDER=2,
ITYPE=3, ISTEP=2, IMORP=1, Q02=4.0, Q2=400.0, DLAM=0.255,
NF=4, XX=0.045, NX=1000, NT=50, NSTEP=100, NXMIN
=$-$4,
NA=1, RFM=0.0, and DKHT=0.0.
The LO evolution is also calculated by choosing IORDER=1.
The results are shown in Fig. 4, in which
the solid (dashed) curves indicates the NLO (LO) calculations
at $x$=0.045, 0.225, and 0.55.
The LO results are different from the RK-MK-SK results
in Ref. \cite{LAG} because the initial distribution is different.
The LO and NLO $xF_3$ are already different at the starting
point ($Q^2$=4 GeV$^2$) due to the NLO contributions from the
coefficient function.
The figure shows that the BF1 evolution agrees
with the experimental $Q^2$ dependence.
As it is obvious from the figure,
the NLO contributions become significant at small $Q^2$
($\approx$1 GeV$^2$).

Next, evolution of the proton $F_2$ structure function
is studied.
As it is shown in the previous subsection, the BF1 program
can be run at small $x$, as small as $x=10^{-4}$,
with good accuracy, so that evolution results could
be compared with recent HERA $F_2$ data.
The initial distributions are the MRS(G) distributions
at $Q_0^2$=4 GeV$^2$.
First, we calculate $F_{2,d^+}(x,Q^2)$ and $F_{2,S}(x,Q^2)$
by setting the initial ones as
QI0=+$0.2513x^{0.335}$$(1+8.63\sqrt{x}
+0.32x)$$(1-x)^{4.41}$+$0.4(1-0.02)$$1.74 x^{-0.067}$$(1-3.45\sqrt{x}
+10.3x)$$(1-x)^{10.1}$+$0.043x^{0.3}(1-x)^{10.1}(1+64.9x)$,
$xq_{_S}$=$2.704x^{0.593}$$(1-0.76\sqrt{x}
+4.20x)$$(1-x)^{3.96}$+$0.2513x^{0.335}$$(1+8.63\sqrt{x}
+0.32x)$$(1-x)^{4.41}$+$1.74 x^{-0.067}$$(1-3.45\sqrt{x}
+10.3x)$$(1-x)^{10.1}$, and
$xg$=$1.51x^{-0.301}$$(1-4.14\sqrt{x}+10.1x)$$(1-x)^{6.06}$
with the input parameters:
IOUT=6, INPUT=1, IREAD=1, MODIFY=1, INDIST=1, IORDER=2,
ITYPE=2, ISTEP=2, IMORP=2, Q02=4.0, Q2=200.0, DLAM=0.255,
NF=4, XX=0.0013, NX
=1000, NT=200, NSTEP=100, NXMIN=$-$4,
NA=1, RFM=0.0, and DKHT=0.0.
The scale parameter is $\Lambda$=0.255 GeV and the number
of flavor is four in the evolution.
Next, $F_{2,s^+}(x,Q^2)$ is calculated with
$xs^+=$$0.2(1-0.02)$$1.74 x^{-0.067}$$(1
-3.45\sqrt{x}+10.3x)$$(1-x)^{10.1}$.
These evolution results
are added to give
$F_2^p(x,Q^2)=-(1/3)[F_{2,d^+}(x,Q^2)$$+F_{2,s^+}$\\
$(x,Q^2)]$$+(4/9)F_{2,S}(x,Q^2)$.
The LO Altarelli-Parisi and Mueller-Qiu evolution results
are also obtained by changing IORDER and MODIFY.
These results are shown in Fig. 5 with various $F_2^p$ data
at $x\approx$0.0013, 0.013, 0.13, and 0.45
from SLAC \cite{SLAC,RWDATA} ($x$=0.14, 0.45),
BCDMS \cite{BCDMS} ($x$=0.14, 0.45),
EMC \cite{EMC,RWDATA} ($x$=0.125, 0.45),
NMC \cite{NMC}($x$=0.0125, 0.14),
Fermilab-E665 \cite{E665} ($x$=0.0012, 0.012),
ZEUS \cite{ZEUS} ($x$=0.0016, 0.014, 0.11),
and H1 \cite {H1} ($x$=0.0013, 0.013, 0.13).
The solid, dashed, dot-dashed curves are obtained
in the LO Altarelli-Parisi, NLO Altarelli-Parisi,
and NLO Mueller-Qiu evolution equations respectively.
The NLO $Q^2$ dependence agrees with the experimental data
as shown in the figure, except for the Fermilab-E665
at small $x$ and at small $Q^2$ where perturbative QCD
would not work.
At this stage, it seems that
$F_2^p$ data are not accurate enough to find
the recombination effects on the evolution.
We do not step into the details of recombination analysis
in comparison with the HERA data.

Finally, $Q^2$ evolution of a nuclear structure function
$F_2^A$ is studied. $Q^2$ dependence of
the ratio $R_2^{Ca}\equiv F_2^{Ca}/F_2^D$
had been measured by NMC \cite{NMCA}.
We compare our evolution with the NMC data in Figs. 6a, 6b, 6c, and 6d
at $x$=0.0085, 0.035, 0.125, and 0.45.
We assume that $F_2^D$ is equal to the nucleon one
for simplicity. The initial distributions at $Q_0^2$=0.8 GeV$^2$
in the nucleon are taken from Ref. \cite{SKF2}:
$xq_{_{NS}}$$=xu_v+xd_v$=$1.967x^{0.501}$$(1-x)^{3.89}$$(1+9.27x)$,
$xq_{_S}$=$1.967x^{0.501}$$(1-x)^{3.89}$$(1+9.27x)$+$2.058
x^{0.294}(1-x)^{11.2}$$(1+7.95x)$, and
$xg$=$16.11x^{0.839}(1-x)^{5.29}$$(1-0.597x)$.
Those in the calcium nucleus are
$xq_{_{NS}}^{Ca}$=$1.840x^{0.472}$$(1-0.984x)^{4.06}$$(1+9.33x)$,
$xq_{_S}^{Ca}$=$1.840x^{0.472}$$(1-0.984x)^{4.06}$$(1
+9.33x)$+$6.423x^{0.600}$$(1-x)^{8.13}$$(1-0.568x)$, and
$xg^{Ca}$=$179.2x^{1.95}$$(1-x)^{7.32}$$(1-0.619x)$.
Both nonsinglet and singlet structure functions are calculated.
Then, we obtain $F_2^A=$$(1/18)F_{2,NS}^A+$$(2/9)F_{2,S}^A$
for isoscalar nucleus because $SU(3)_{flavor}$ symmetry
is assumed in antiquark distributions in Ref. \cite{SKF2}.
In Figs. 6a, 6b, 6c, and 6d,
the solid, dashed, dot-dashed curves are obtained
in the LO Altarelli-Parisi, NLO Altarelli-Parisi,
and NLO Mueller-Qiu evolution equations respectively
with $\Lambda$=0.2 GeV and $N_f$=3.
As shown by the figures, NLO and recombination
contributions to the ratio
are conspicuous at small $x$ (=0.0085, 0.035).
They are very small at medium $x$ (=0.45) in Fig. 6d; however,
it does not mean that their contributions to
the structure functions themselves are very small.
If we evolve $F_2$ from $Q_0^2$=0.8 GeV$^2$,
the recombination effects are larger than the NLO ones.
It is interesting to find such large recombination
contributions in Fig. 6a.
However, the recombination cannot be tested at this stage
because we do not have
the data in the wide $Q^2$ region at small $x$.

\vfill\eject
\vspace{1.0cm}
\noindent
{\Large\bf {7. Conclusions}}
\vspace{0.4cm}

We investigated numerical solution of $Q^2$ evolution equations
of spin-independent parton distributions (and structure functions)
in the nucleon and in nuclei by using the brute-force method.
Two types of evolution equations are studied. One is the usual
Altarelli-Parisi equations and the other is the modified ones
due to parton recombinations (Mueller-Qiu).

We divide the variables $x$ and $t$ into small steps and
simply solve the evolution equations.
Numerical accuracy depends essentially on the numbers $N_x$ and $N_t$.
We obtain excellent numerical solution in the wide
$x$ range ($0.0001\le x\le 0.8$) by taking
large enough $N_x$ and $N_t$.
We recommend to use $N_x\ge 1000$ and $N_t\ge 50$
in the nonsinglet case and $N_x\ge 1000$ and $N_t\ge 200$
in the singlet for getting numerical accuracy better
than 2\% in the $x$ range, $0.0001<x<0.8$.

We provide the FORTRAN program BF1, which can be run
by supplying the input parameters and the input distribution(s).
Parton distributions $xq_{_{NS}}$, $xq_{_{S}}$, $xq_i+x\bar q_i$
(corresponding structure functions), $xg$, and those
in nuclei could be evolved in the program.
The program can also handle the $Q^2$ devolution from $Q_0^2$
to smaller $Q^2$.
This is a very useful program in studying
structure functions in the nucleon and in nuclei.
Typical running CPU time is five minutes on the AlphaServer 2100 4/200
in the nonsinglet and one hour in the singlet.
Therefore, a reasonably powerful machine should be used for running
the BF1 program in the singlet case.

\vspace{1.0cm}
\noindent
{\Large\bf Acknowledgments}
\vspace{0.4cm}

This research was partly supported by the Grant-in-Aid for
Scientific Research from the Japanese Ministry of Education,
Science, and Culture under the contract number 06640406
and also by the U.S. Department of Energy under the contract number
DE-AC02-76CH00016.
MM and SK thank the Research Center for Nuclear Physics
in Osaka for making them use computer facilities.
They thank W. Melnitchouk for communications on
NLO effects in Ref. \cite{WALLY} and thank
A. Br\"ull, E. Rondio, and H. M. Schellman for providing them
experimental data.
SK was supported by the Japan Society for Promotion of Science
through the Japan/US cooperation program
in the field of high-energy physics for his visiting
Brookhaven National Laboratory (BNL).
They would like to thank BNL for its hospitality during their stay.

\vfill\eject
\noindent
{\Large\bf {Appendix A. Running coupling constants}}
\vspace{0.4cm}

Running coupling constant in the leading order (LO) is
$$
\alpha_s^{LO}(Q^2) = {4\pi \over \beta_0 \ln (Q^2/\Lambda^2)}
\ \ \ ,
\eqno{(A.1)}
$$
and the one in the next-to-leading order (NLO) is \cite{RGR}
$$
\alpha_s^{NLO}(Q^2)={4\pi \over \beta_0 \ln(Q^2/\Lambda^2)}
\Biggl[1-{\beta_1 \ln \ln(Q^2/\Lambda^2) \over \beta_0^2\ln(Q^2/\Lambda^2)}
\Biggr]
\ \ \ .
\eqno{(A.2)}
$$
In Eq. (A.2), the renormalization scheme is $\overline{MS}$ and
$\Lambda$ is the QCD scale parameter in this scheme.
$\beta_0$ and $\beta_1$ are given by
$$
\beta_0={11\over3}C_G-{4 \over 3}T_R N_f  \ \ \ , \ \ \
\beta_1={34\over3}C^2_G-{10\over3}C_G N_f-2C_F N_f
\ \ \ ,
\eqno{(A.3)}
$$
with the color constants
$$
C_G=N_c~~~,~~~ C_F={N_c^2-1 \over 2N_c}~~~,~~~ T_R={1 \over2} ~~~.
\eqno{(A.4)}
$$
$N_c$ is the number of color ($N_c$=3) and
$N_f$ is the number of flavor.

\vspace{1.0cm}
\noindent
{\Large\bf {Appendix B. Splitting functions}}
\vspace{0.4cm}

Splitting functions in the leading order are
\begin{eqnarray*}
P_{qq}^{(0)}(x) &=&  C_F\ \left[\ {{1+x^2} \over {(1-x)_+}}
                    \ + \ {3 \over 2} \ \delta (1-x)\ \right]\ \ \ , \\
P_{qg}^{(0)}(x) &=& 2T_R \left[\ x^2 \ + \ (1-x)^2\ \right]\ \ \ , \\
P_{gq}^{(0)}(x) &=& C_F \ {{1+(1-x)^2} \over x} \ \ \ , \\
P_{gg}^{(0)}(x) &=& 2C_G \left[\ {x \over {(1-x)_+}} \ + \
                        {{1-x} \over x} \ + \ x(1-x) \ + \
           \left( {{11}\over{12}} - {1 \over 3}{{N_fT_R} \over {C_G}}\right)
           \ \delta (1-x) \ \right]\ \ \ , \\
\end{eqnarray*}
\vspace{-3.5cm}

\hfill{(B.1)}

\vspace{+2.0cm}
\noindent
where the $+$ function is defined by
$$
\int_0^1 dx {{f(x)} \over {(1-x)_+}} \ = \
\int_0^1 dx {{f(x)-f(1)} \over {1-x}}
\ \ \ .
\eqno{(B.2)}
$$
It should be noted that the above integration is defined
in the region $0\le x\le 1$.

The NLO splitting functions for $q_i^+ (x)$ and $g(x)$ are
given by Herrod and Wada
\cite{HWW,OTHERAP}
\begin{eqnarray*}
P_{q_i^+ q_j^+}^{(1)}(x) &=&  \delta_{ij} \ P_{q^+,NS}^{(1)}(x) \ + \
                              2 C_F T_R F_{qq}(x) \ \ \ , \\
P_{q_i^+ g}^{(1)}(x) &=& - \ C_G T_R F_{qg}^1(x)
                       \ - \ C_F T_R F_{qg}^2(x) \ \ \ , \\
P_{gq_i^+}^{(1)}(x) &=& - \ C_F^2 F_{gq}^1(x)
                      \ - \ C_F C_G F_{gq}^2(x)
                      \ - \ C_F T_R N_f F_{gq}^3(x) \ \ \ , \\
P_{gg}^{(1)}(x) &=&      C_G T_R N_f F_{gg}^1(x)
                   \ + \ C_F T_R N_f F_{gg}^2(x)
                   \ + \ C_G^2 F_{gg}^3(x) \ \ \ . \\
\end{eqnarray*}
\vspace{-3.5cm}

\hfill{(B.3)}

\vspace{+2.0cm}
\noindent
The NLO nonsinglet splitting function for a ``$q \pm \bar q$ type"
distribution [$\displaystyle{q_{_{NS}}=\sum_i a_i (q_i \pm \bar q_i)}$]
is given by \cite{HWW}
\begin{eqnarray*}
P_{q^\pm, NS}^{(1)}(x)&=&C_F^2\biggl\{P_F(x) \pm P_A(x)+\delta(1-x)
\biggl[{3\over8}-{1\over2}\pi^2
+\zeta(3)-8\widetilde{S}(\infty)\biggr]\biggr\}  \\
&+&{1\over2}C_FC_A\bigg\{P_G(x) \mp P_A(x)+\delta(1-x)
\biggl[{17\over12}+{11\over9}
\pi^2-\zeta(3)+8\widetilde{S}(\infty)\biggr]\biggr\}  \\
&+&C_FT_RN_f\biggl\{P_{N_F}(x)-\delta(1-x)
\biggl({1\over6}+{2\over9}\pi^2\biggr)
\biggr\}
\end{eqnarray*}
\vspace{-1.5cm}

\hfill{(B.4)}

\vspace{+0.5cm}
\noindent
where $P_F(x)$, $P_G(x)$, $P_{N_F}(x)$, and $P_A(x)$ are given
by Curci, Furmanski, and Petronzio \cite{OTHERAP} as
\begin{eqnarray*}
P_F(x)\!\!&=&\!\!-2{{1+x^2} \over {1-x}}\ln x\ln(1-x)-
\biggl({3\over {1-x}}+2x\biggr)\ln x-{1\over2}(1+x)\ln^2x  -\!\!5(1-x)   ~,\\
P_G(x)\!\!&=&\!\!{1+x^2\over (1-x)_+}\bigg[\ln^2x+{11 \over3}\ln x+{67\over9}
-{1\over3}\pi^2\biggr]+2(1+x)\ln x+{40\over 3}(1-x)                     ~~~, \\
P_{N_F}(x)\!\!&=&\!\!{2\over 3}\Biggl[{1+x^2\over (1-x)_+}\biggl(
-\ln x-{5\over 3}\biggr)-2(1-x)\Biggr]                                  ~~~, \\
P_A(x)\!\!&=&\!\!2{1+x^2\over 1+x}\int_{x/(1+x)}^{1/(1+x)}{dz \over z}
\ln {{1-z} \over z}
+2(1+x)\ln x+4(1-x)                                                     ~~~.
\end{eqnarray*}
\vspace{-1.5cm}

\hfill{(B.5)}

\vspace{+0.5cm}
\noindent
The $\zeta$ function is defined by
$\displaystyle{\zeta (k)=\sum_{n=1}^\infty {1 \over n^k}}$
and the numerical value ($\zeta(3)$=1.2020569...)
is taken from Ref. \cite{AS}. $\widetilde S(\infty)$ is given
by the $\zeta$ function as
$ \widetilde S (\infty) =-{5 \over 8}\zeta(3) $.
The integral in $P_A(x)$ is expressed in terms of
the Spence function $S(x)$:
$$
\int_{x/(1+x)}^{1/(1+x)}{dz \over z} \ln {{1-z} \over z} =
S\left({x\over{1+x}}\right)  -
S\left({1\over{1+x}}\right)  -
{1\over 2}  \left[\ \ln^2 {1\over{1+x}}  -
                     \ln^2 {x\over{1+x}} \ \right]
\ \ \ ,
\eqno{(B.6)}
$$
where $S(x)$ is defined by
$$
S(x) \ = \ \int_x^1 dz \ {{\ln z}\over{1-z}}
\ \ \ .
\eqno{(B.7)}
$$
It should be noted that other convention is sometimes used,
namely $-S(x)$ may be called the Spence function \cite{AS}.
It is useful to use a series expansion form for numerical
calculations \cite{AS}:
$$
S(x) \ = \ - \ \sum_{k=1}^\infty {{(1-x)^k} \over {k^2}}
\ \ \ .
\eqno{(B.8)}
$$

The functions $F_{qq}$, $F_{qg}$, $F_{gq}$, and $F_{gg}$ in Eq. (B.3)
are defined by
\begin{eqnarray*}
{F}_{qq}(x)&=&{\frac{20}{9}}{\frac{1}{x}}-2+6x
-{\frac{56}{9}}{x}^{2}+\left({1+5x+{\frac{24}{9}}{x}^{2}}\right)
{\ln}x-\left({1+x}\right){\ln}^{2}x \ \ \ , \\
{F}_{qg}^{1}(x)&=&-{\frac{40}{9}}{\frac{1}{x}}+4-50x+{\frac{436}{9}}{x}^{2}
-\left({2+16x+{\frac{88}{3}}{x}^{2}}\right){\ln}x\\
& &+(2+4x){\ln}^{2}x+8x\left({1-x}\right)\ln\left({1-x}\right)\\
& &+\left({2-4x+4{x}^{2}}\right)\left[{{\ln}^{2}\left({1-x}\right)
-{\frac{1}{6}}{\pi }^{2}}\right]
-4\left({1+2x+2{x}^{2}}\right){J}_{x} \ \ \ , \\
{F}_{qg}^{2}(x)&=&-14+29x-20{x}^{2}+\left({-3+4x-8{x}^{2}}\right){\ln}x\\
& &-\left({1-2x+4{x}^{2}}\right){\ln}^{2}x
-8x\left({1-x}\right)\ln\left({1-x}\right)\\
& &+\left({2-4x+4{x}^{2}}\right)\left[{-{\ln}^{2}\left({1-x}\right)
+2{\ln}x\ln\left({1-x}\right)+{\frac{1}{3}}{\pi }^{2}}\right] \ \ \ , \\
{F}_{gq}^{1}(x)&=&{\frac{5}{2}}+{\frac{7}{2}}x
-\left({2+{\frac{7}{2}}x}\right){\ln}x
+\left({1-{\frac{1}{2}}x}\right){\ln}^{2}x\\
& &+\left({{\frac{6}{x}}-6+5x}\right)\ln\left({1-x}\right)
+\left({{\frac{2}{x}}-2+x}\right){\ln}^{2}\ \left({1-x}\right)
\ \ \ , \\
{F}_{gq}^{2}(x)&=&-{\frac{1}{x}}-{\frac{19}{9}}
-{\frac{37}{9}}x-{\frac{44}{9}}{x}^{2}
+\left({12+5x+{\frac{8}{3}}{x}^{2}}\right){\ln}x \\
& &-\left({2+x}\right){\ln}^{2}x
-{\frac{1}{3}}\left({{\frac{22}{x}}-22+17x}\right)\ln\left({1-x}\right)\\
& &-\left({{\frac{2}{x}}-2+x}\right)\left[{{\ln}^{2}\left({1-x}\right)
-2{\ln}x\ln\left({1-x}\right)-{\frac{1}{6}}{\pi }^{2}}\right]\\
& &+\left({{\frac{4}{x}}+4+2x}\right)\ {J}_{x} \ \ \ , \\
{F}_{gq}^{3}(x)&=&{\frac{40}{9}}{\frac{1}{x}}-{\frac{40}{9}}
+{\frac{32}{9}}x
+{\frac{4}{3}}\left({{\frac{2}{x}}-2+x}\right)\ln\left({1-x}\right)
\ \ \ , \\
{F}_{gg}^{1}(x)&=&-{\frac{20}{9}}{\frac{1}{{\left({1-x}\right)}_{+}}}
-{\frac{2}{9}}\left({{\frac{23}{x}}-29+19x-23{x}^{2}}\right)\\
& &-{\frac{4}{3}}\left({1+x}\right){\ln}x
-{\frac{4}{3}}\delta\left({1-x}\right) \ \ \ , \\
\end{eqnarray*}

\vfill\eject

\begin{eqnarray*}
{F}_{gg}^{2}(x)&=&{\frac{4}{3x}}-16+8x+{\frac{20}{3}}{x}^{2}
-\left({6+10x}\right){\ln}x\\
& &-2\left({1+x}\right){\ln}^{2}x-\delta\left({1-x}\right) \ \ \ , \\
{F}_{gg}^{3}(x)&=&{\frac{67}{9}}{\frac{1}{{\left({1-x}\right)}_{+}}}
-{\frac{8}{9}}-{\frac{1}{18}}x-\left({{\frac{47}{6}}
-{\frac{25}{6}}x+{\frac{44}{3}}{x}^{2}}\right){\ln}x\\
& &+\left({{\frac{1}{1-x}}+{\frac{1}{1+x}}
+{\frac{5}{4}}+6x-2{x}^{2}}\right){\ln}^{2}x\\
& &+\left({-{\frac{11}{2x}}+{\frac{7}{2}}
+{\frac{13}{4}}x}\right)\ln\left({1-x}\right)
+{\frac{1}{4}}\left({-{\frac{5}{x}}+5+2x}\right)
{\ln}^{2}\left({1-x}\right)\\
& &-\left({{\frac{4}{1-x}}+{\frac{4}{x}}-5+7x-4{x}^{2}}\right)
{\ln}x\ln\left({1-x}\right)+\left({1+x}\right){I}_{x}\\
& &+4\left({{\frac{1}{1+x}}-{\frac{1}{x}}-2-x-{x}^{2}}\right){J}_{x}
-{\frac{1}{3}}{\pi }^{2}\left({{\frac{1}{{\left({1-x}\right)}_{+}}}
+{\frac{1}{x}}-1+2x-{x}^{2}}\right)\\
& &+\delta\left({1-x}\right)\left[{-4\widetilde{S}\left({\infty}\right)
+{\frac{1}{2}}\zeta \left({3}\right)+{\frac{8}{3}}}\right]
\ \ \ .
\end{eqnarray*}
\vspace{-4.5cm}

\hfill{(B.9)}

\vspace{3.8cm}
\noindent
where $I_x$ is the Spence function [$I_x=S(x)$] and $J_x$
is defined by
$$
J_x = -{\frac{1}{12}}{\pi}^2 - \ln x\ln (1+x) + S(1+x)
\ \ \ .
\eqno{(B.10)}
$$

In the one-step evolution equation in Eq. (2.14),
the matrix ${\bf E}$ is introduced. Each matrix element
is given by
\begin{eqnarray*}
{E}_{qq}\left({x}\right)&=&-{E}\left({{\gamma}_{gq}{C}^{g}}\right)
\ \ \ , \\
{E}_{qg}\left({x}\right)&=&{E}\left({{\gamma}_{qq}{C}^{g}}\right)
-{E}\left({{\gamma}_{qg}{C}^{q}}\right)
-{E}\left({{\gamma}_{gg}{C}^{g}}\right) \ \ \ , \\
{E}_{gq}\left({x}\right)&=&{E}\left({{\gamma}_{gq}{C}^{q}}\right)
\ \ \ , \\
{E}_{gg}\left({x}\right)&=&-{E}_{qq}\left({x}\right) \ \ \ .
\end{eqnarray*}
\vspace{-2.5cm}

\hfill{(B.11)}

\vspace{+1.5cm}
\noindent
where
\begin{eqnarray*}
{E}\left({{\gamma}_{gq}{C}^{g}}\right)/2{C}_{F}{T}_{R}&=&-{\frac{2}{3x}}
+{\frac{20}{3}}-{\frac{2x}{3}}
-{\frac{16{x}^{2}}{3}}+\left({1+5x-{\frac{4}{3}}{x}^{2}}\right){\ln}x\\
& &+\left({1+x}\right)\left({{\ln}^{2}\ x-2{\ln}x\ln\left({1-x}\right)
+2{I}_{x}}\right)\\
& &+\left({-{\frac{4}{3x}}-1+x
+{\frac{4}{3}}{x}^{2}}\right)\ln\left({1-x}\right) \ \ \ , \\
\end{eqnarray*}

\vfill\eject
\begin{eqnarray*}
{E}\left({{\gamma}_{qq}{C}^{g}}\right)/{C}_{F}{T}_{R}&=&-7+10x
-\left({1-16x+32{x}^{2}}\right){\ln}x
-\left({1-2x+4{x}^{2}}\right){\ln}^{2}x\\
& &+\left({6-12x+16{x}^{2}}\right){\ln}x\ln\left({1-x}\right)\\
& &+\left({5-36x+32{x}^{2}}\right)\ln\left({1-x}\right)\\
& &-4\left({1-2x+2{x}^{2}}\right)\left[{{\ln}^{2}\left({1-x}\right)
-{\frac{{\pi }^{2}}{6}}}\right]-2\left({1-2x+4{x}^{2}}\right){I}_{x}
\ \ \ , \\
{E}\left({{\gamma}_{qg}{C}^{q}}\right)/{C}_{F}{T}_{R}&=&5-12x+16{x}^{2}
+\left({1+8x-12{x}^{2}}\right){\ln}x\\
& &+\left({1-2x+4{x}^{2}}\right)\left[{-{\ln}^{2}x
+2{\ln}x\ln\left({1-x}\right)}\right]\\
& &+\left({7-16x+12{x}^{2}}\right)\ln\left({1-x}\right)\\
& &+\left({1-2x+2{x}^{2}}\right)\left[{-2{\ln}^{2}\left({1-x}\right)
+{\frac{2}{3}}{\pi }^{2}}\right]+2\left({1-2x}\right){I}_{x} \ \ \ , \\
{E}\left({{\gamma}_{gq}{C}^{q}}\right)
/{\frac{1}{2}}{C}_{F}^{2}&=&3+6x
-3x{\ln}x+\left({2-x}\right){\ln}^{2}x\\
& &+\left({{\frac{12}{x}}-16+7x}\right)\ln\left({1-x}\right)\\
& &+2\left({{\frac{2}{x}}-2+x}\right)\left[{-{\ln}^{2}\left({1-x}\right)
+{\frac{1}{3}}{\pi }^{2}}\right]\\
& &-2\left({2-x}\right){\ln}x\ln\left({1-x}\right)
+2\left({{\frac{4}{x}}-2+x}\right){I}_{x} \ \ \ , \\
{E}\left({{\gamma}_{gg}{C}^{g}}\right)&=&{C}_{G}{T}_{R}A
+{\frac{4}{3}}{T}_{R}^{2}{N}_{f}B \ \ \ .
\end{eqnarray*}
\vspace{-3.5cm}

\hfill{(B.12)}

\vspace{+2.8cm}
\noindent
$A$ and $B$ in the above equation are
\begin{eqnarray*}
A&=&-{\frac{4}{3x}}+18+{\frac{154}{3}}x-{\frac{193}{3}}{x}^{2}
+\left({{\frac{17}{3}}+{\frac{170}{3}}x
-{\frac{40}{3}}{x}^{2}}\right){\ln}x\\
& &+2\left({1+4x}\right)\left({{\ln}^{2}x
+2{I}_{x}}\right)-\left({{\frac{8}{3x}}+{\frac{5}{3}}+{\frac{122}{3}}x
-{\frac{136}{3}}{x}^{2}}\right)\ln\left({1-x}\right)\\
& &-4\left({1-2x+2{x}^{2}}\right)\left[{{\ln}^{2}\left({1-x}\right)
-{\frac{{\pi }^{2}}{6}}}\right]
-8x\left({3-x}\right){\ln}x\ln\left({1-x}\right) \ \ \ , \\
B&=&-1+8x-8{x}^{2}
-\left({1-2x+2{x}^{2}}\right)\ln\left({{\frac{x}{1-x}}}\right)
\ \ \ .
\end{eqnarray*}
\vspace{-1.5cm}

\hfill{(B.13)}

\vspace{0.5cm}
\noindent
In the above equation, we corrected the factor
$(17/3+170x/3-40x^2/3)$, which was
$(17+170x/3-40x^2/3)$ in the original paper \cite{HWW}.

The splitting function $\widetilde {\bar P}_{qg}=x \bar P_{qg}$
is given by
$$
\bar {P}_{qg}(x) = 2 ~(-2x+15x^2-30x^3+18x^4) \ \ \ .
\eqno{(B.14)}
$$

\vfill\eject
\noindent
{\Large\bf {Appendix C. Coefficient functions}}
\vspace{0.4cm}

The coefficient functions $C_n^{q,g}(x,\alpha_s)$ in Eqs. (2.10) and
(2.11) are written by the functions $B_n^{q,g}(x)$:
$$
C_n^{q}(x,\alpha_s)  =  \delta(1-x)  +
                {{\alpha_s}\over{4\pi}} B_n^{q}(x)
\ \ \ \ , \ \ \ \
C_n^{g}(x,\alpha_s)  =
                {{\alpha_s}\over{4\pi}} B_n^{g}(x)
\ \ \ .
\eqno{(C.1)}
$$
The quark part is given by \cite{HWW}
\begin{eqnarray*}
B_1^{q}(x)&=&{C_F \over 2}[F_q(x)-4x]  \ \ \ , \\
B_2^{q}(x)&=&C_F F_q(x)                \ \ \ , \\
B_3^{q}(x)&=&C_F [F_q(x)-2-2x]         \ \ \ ,
\end{eqnarray*}
\vspace{-1.5cm}

\hfill{(C.2)}

\vspace{+0.5cm}
\noindent
where the function $F_q(x)$ is given by
$$
F_q(x)=-{3 \over 2}{1+x^2 \over (1-x)_+}
        +{1\over 2}(9+5x)-2{1+x^2 \over 1-x}
          \ln x +2(1+x^2)\biggl[ {\ln (1-x) \over 1-x} \biggr]_+
                  -\delta (1-x)(9+{2 \over 3}\pi^2) \ \ .
\eqno{(C.3)}
$$
The gluon part is:
\begin{eqnarray*}
B_1^{g}(x)&=&2T_R [F_g(x)- 4x(1-x)]  \ \ \ , \\
B_2^{g}(x)&=&4T_R F_g(x)            \ \ \ ,
\end{eqnarray*}
\vspace{-1.5cm}

\hfill{(C.4)}

\vspace{0.5cm}
\noindent
where
$$
F_g(x)=  (1-2x+2x^2) \ln \biggl( {{1-x} \over x} \biggr)
                           +8x(1-x)-1
\ \ \ .
\eqno{(C.5)}
$$

\vfill\eject

\vfill\eject
\noindent
{\Large\bf{Figure Captions}} \\

\vspace{-0.38cm}
\begin{description}
   \item[Fig. 1]
(a) $N_t$ dependence of nonsinglet evolution results
    is shown. Next-to-leading-order Altarelli-Parisi
    evolution results are calculated by running the program BF1.
    $N_x$=1000 is fixed and $N_t$ is varied ($N_t$=10, 20, and 500)
    in order to check numerical accuracy due to the choice of $N_t$.
    The initial distribution is the HMRS-B $xu_v+xd_v$ distribution
    at $Q_0^2$=4 GeV$^2$.
    There are dotted, dashed, and solid curves at $Q^2$=200 GeV$^2$ for
    $N_t$=10, 20, and 500 respectively; however, they cannot be
    distinguished in the figure because differences among them
    are fairly small. See text for the details of the input parameters.
(b) $N_x$ dependence is shown by taking fixed $N_t$=50.
    The dotted, dashed, dot-dashed, and solid curves are the results
    for $N_x$=100, 300, 1000, and 4000 respectively.
   \item[Fig. 2]
(a) $N_t$ dependence of singlet-quark evolution results
    is shown. Next-to-leading-order Altarelli-Parisi
    evolution results are calculated.
    $N_x$=1000 is fixed and $N_t$ is varied.
    The initial distributions are the HMRS-B $xq_{_S}$ and $xg$
    distributions at $Q_0^2$=4 GeV$^2$.
    The dotted, dashed, dot-dashed, and solid curves
    at $Q^2$=200 GeV$^2$ for $N_t$=20, 50, 200, and 500 respectively.
(b) $N_x$ dependence is shown by taking fixed $N_t$=200.
    The dotted, dashed, dot-dashed, and solid curves are the results
    for $N_x$=100, 300, 1000, and 4000 respectively.
   \item[Fig. 3]
(a) $N_t$ dependence of gluon evolution results
    is shown. $N_x$ is fixed at $N_x$=1000.
    The input parameters and the input distributions
    are the same with those in Fig. 2.
    The dotted, dashed, dot-dashed, and solid curves
    at $Q^2$=200 GeV$^2$ are for $N_t$=20, 50, 200, and 500.
(b) $N_x$ dependence is shown by taking fixed $N_t$=200.
    The dotted, dashed, dot-dashed, and solid curves are the results
    for $N_x$=100, 300, 1000, and 4000 respectively.
   \item[Fig. 4]
Nonsinglet $Q^2$ evolution results are compared with CDHSW data.
The initial distribution is the MRS(G) $xu_v+xd_v$ at $Q_0^2$=4 GeV$^2$.
The solid and dashed curves are the NLO and LO results
at $x$=0.045, 0.225, and 0.55.
   \item[Fig. 5]
$Q^2$ evolution results of the proton's $F_2$ structure function
at $x$=0.0013, 0.013, 0.13, and 0.45
are compared with various data in SLAC, BCDMS, EMC, NMC, Fermilab-E665,
ZEUS, and H1 experiments.
The solid, dotted, and dashed curves are obtained by using
the NLO Altarelli-Parisi, LO Altarelli-Parisi, and
NLO Mueller-Qiu equations respectively.
$F_2^p$ at $x$=0.0013, 0.013, and 0.13 are shown by
$10 \cdot F_2^p(x=0.0013)$, $10^{2/3}\cdot F_2^p(x=0.013)$, and
$10^{1/3}\cdot F_2^p(x=0.13)$.
   \item[Fig. 6]
$Q^2$ dependence of the structure function ratio
$F_2^{Ca}(x,Q^2)/F_2^D(x,Q^2)$ is calculated and results are
compared with NMC data.
The solid, dotted, and dashed curves are obtained by using
the NLO Altarelli-Parisi, LO Altarelli-Parisi, and
NLO Mueller-Qiu equations ($\Lambda$=0.2 GeV, $N_f$=3)
respectively at
(a) x=0.0085, (b) 0.035, (c) 0.125, and (d) 0.45.
The input distributions in the nucleon and in the calcium
nucleus are taken from Ref. \cite{SKF2}.
For example, AP (NLO) means that the next-to-leading-order Altarelli-Parisi
equations are used in evolving $F_2$ in the nucleon and the
one in the calcium, then the ratios are taken.
\end{description}

\vfill\eject
\noindent
{\Large\bf{TEST RUN OUTPUT}} \\
\vspace{0.4cm}

\noindent
IOUT= 3\ \ \ INPUT= 1\ \ \ IREAD= 1\ \ \ MODIFY= 1\ \ \ INDIST= 1 \\
IORDER= 2\ \ \ ITYPE= 4\ \ \ ISTEP= 1\ \ \ IMORP= 2 \\
Q02= 4.0000\ \ \ Q2= 200.000\ \ \ DLAM= 0.2550\ \ \ NF= 4 \\
XX= 0.0000000\ \ \ NX=1000\ \ \ NT= 200\ \ \ NSTEP=  50\ \ \ NXMIN= $-$4 \\
NA=   1\ \ \ RFM= 0.00\ \ \ DKHT= 0.000 \\

\begin{tabbing}
     0.000100 \ \ \ \ \= 18.335139 \ \ \ \ \= 66.652696 \=  \kill
     0.000100 \> 18.335139  \> 66.652696 \> \\
     0.000120 \> 17.128091  \> 61.663113 \> \\
     0.000145 \> 15.996575  \> 57.000809 \> \\
     0.000174 \> 14.936139  \> 52.646335 \> \\
     0.000209 \> 13.942574  \> 48.581338 \> \\
     0.000251 \> 13.011907  \> 44.788502 \> \\
     0.000302 \> 12.140385  \> 41.251490 \> \\
     0.000363 \> 11.324464  \> 37.954894 \> \\
     0.000437 \> 10.560801  \> 34.884181 \> \\
     0.000525 \> \ 9.846243 \> 32.025646 \> \\
     0.000631 \> \ 9.177817 \> 29.366368 \> \\
     0.000759 \> \ 8.552722 \> 26.894167 \> \\
     0.000912 \> \ 7.968325 \> 24.597561 \> \\
     0.001096 \> \ 7.422150 \> 22.465735 \> \\
     0.001318 \> \ 6.911875 \> 20.488499 \> \\
     0.001585 \> \ 6.435325 \> 18.656257 \> \\
     0.001905 \> \ 5.990469 \> 16.959975 \> \\
     0.002291 \> \ 5.575416 \> 15.391153 \> \\
     0.002754 \> \ 5.188411 \> 13.941795 \> \\
     0.003311 \> \ 4.827834 \> 12.604385 \> \\
     0.003981 \> \ 4.492194 \> 11.371860 \> \\
     0.004786 \> \ 4.180129 \> 10.237586 \> \\
     0.005754 \> \ 3.890402 \> \ 9.195334 \> \\
     0.006918 \> \ 3.621892 \> \ 8.239259 \> \\
     0.008318 \> \ 3.373589 \> \ 7.363868 \> \\
     0.010000 \> \ 3.144582 \> \ 6.564002 \> \\
     0.012023 \> \ 2.934037 \> \ 5.834797 \> \\
     0.014454 \> \ 2.741170 \> \ 5.171655 \> \\
     0.017378 \> \ 2.565211 \> \ 4.570205 \> \\
     0.020893 \> \ 2.405346 \> \ 4.026253 \> \\
     0.025119 \> \ 2.260649 \> \ 3.535733 \> \\
     0.030200 \> \ 2.129987 \> \ 3.094648 \> \\
     0.036308 \> \ 2.011924 \> \ 2.699013 \> \\
     0.043652 \> \ 1.904595 \> \ 2.344797 \> \\
     0.052481 \> \ 1.805604 \> \ 2.027887 \> \\
     0.063096 \> \ 1.711949 \> \ 1.744079 \> \\
     0.075858 \> \ 1.620025 \> \ 1.489125 \> \\
     0.091201 \> \ 1.525750 \> \ 1.258859 \> \\
     0.109648 \> \ 1.424866 \> \ 1.049428 \> \\
     0.131826 \> \ 1.313447 \> \ 0.857641 \> \\
     0.158489 \> \ 1.188551 \> \ 0.681417 \> \\
     0.190546 \> \ 1.048903 \> \ 0.520250 \> \\
     0.229087 \> \ 0.895387 \> \ 0.375536 \> \\
     0.275423 \> \ 0.731157 \> \ 0.250490 \> \\
     0.331131 \> \ 0.561477 \> \ 0.149345 \> \\
     0.398107 \> \ 0.393883 \> \ 0.075655 \> \\
     0.478630 \> \ 0.239332 \> \ 0.029963 \> \\
     0.575440 \> \ 0.113441 \> \ 0.007968 \> \\
     0.691831 \> \ 0.033130 \> \ 0.001021 \> \\
     0.831764 \> \ 0.002843 \> \ 0.000023 \> \\
     1.000000 \> \ 0.000000 \> \ 0.000000 \> \\
\end{tabbing}

\vfill\eject

\hspace{2.7cm}
\epsfysize=10.5cm
\epsfbox{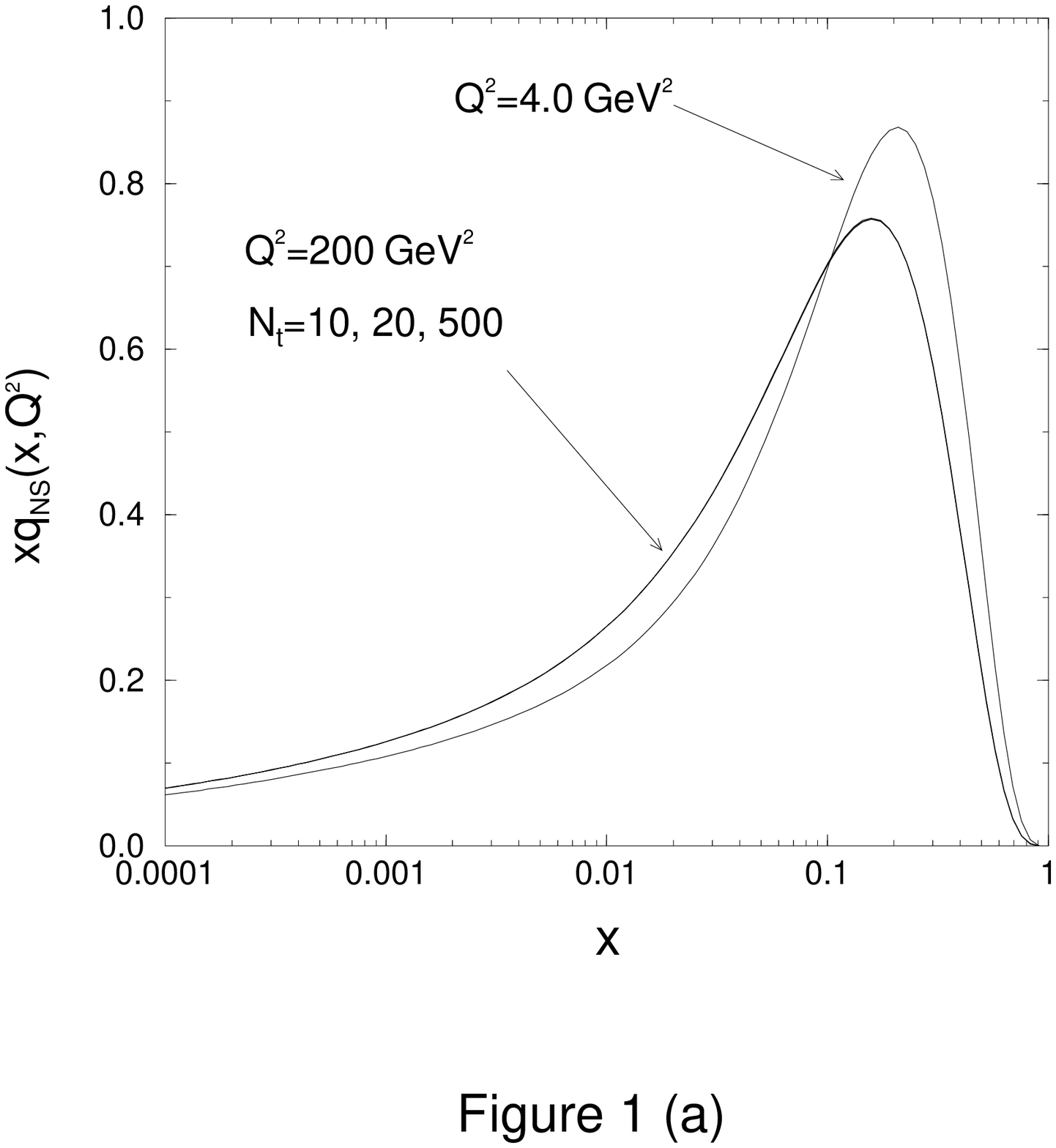}

\hspace{2.7cm}
\epsfysize=10.5cm
\epsfbox{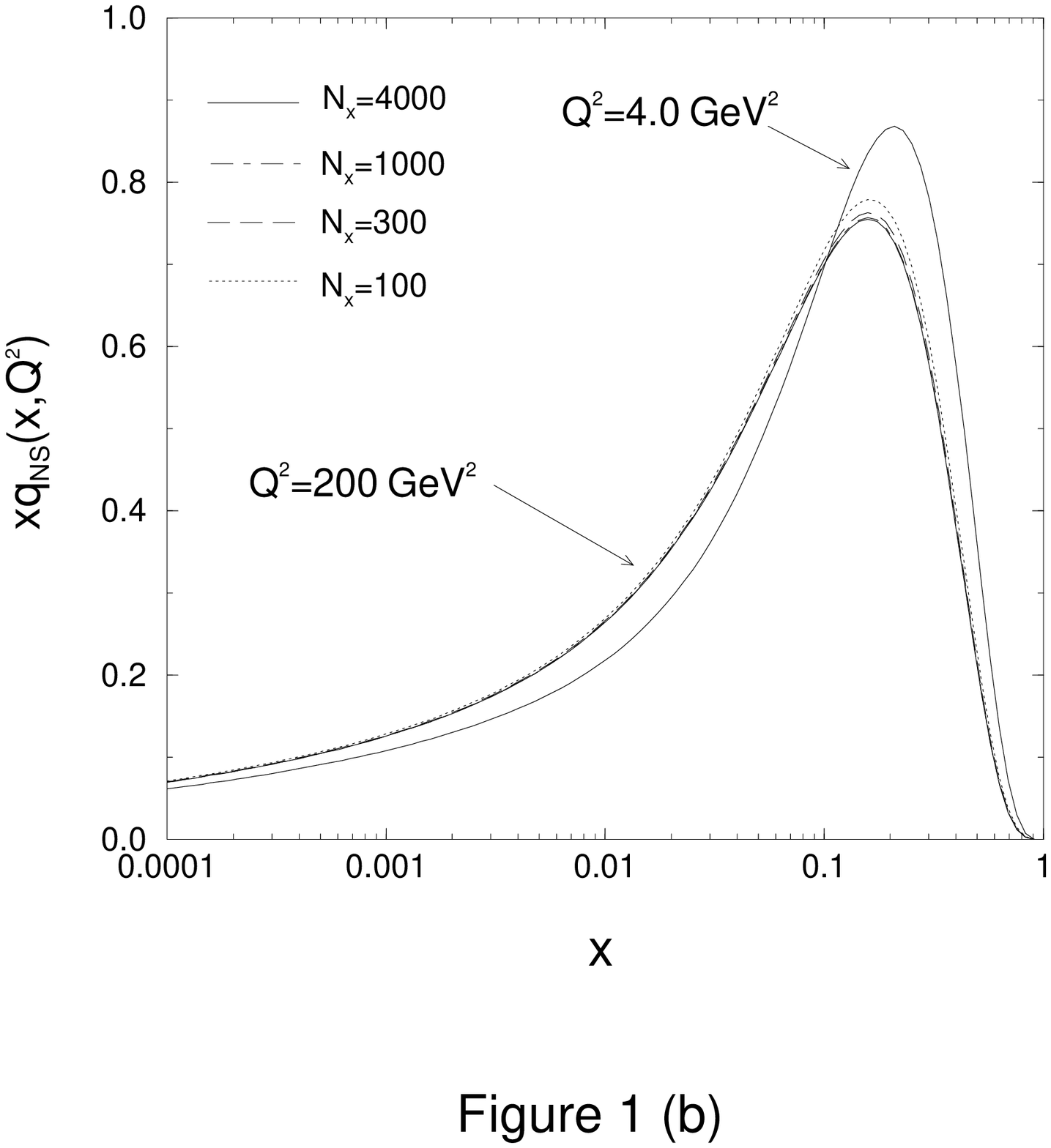}
\vfill\eject
\hspace{2.7cm}
\epsfysize=10.5cm
\epsfbox{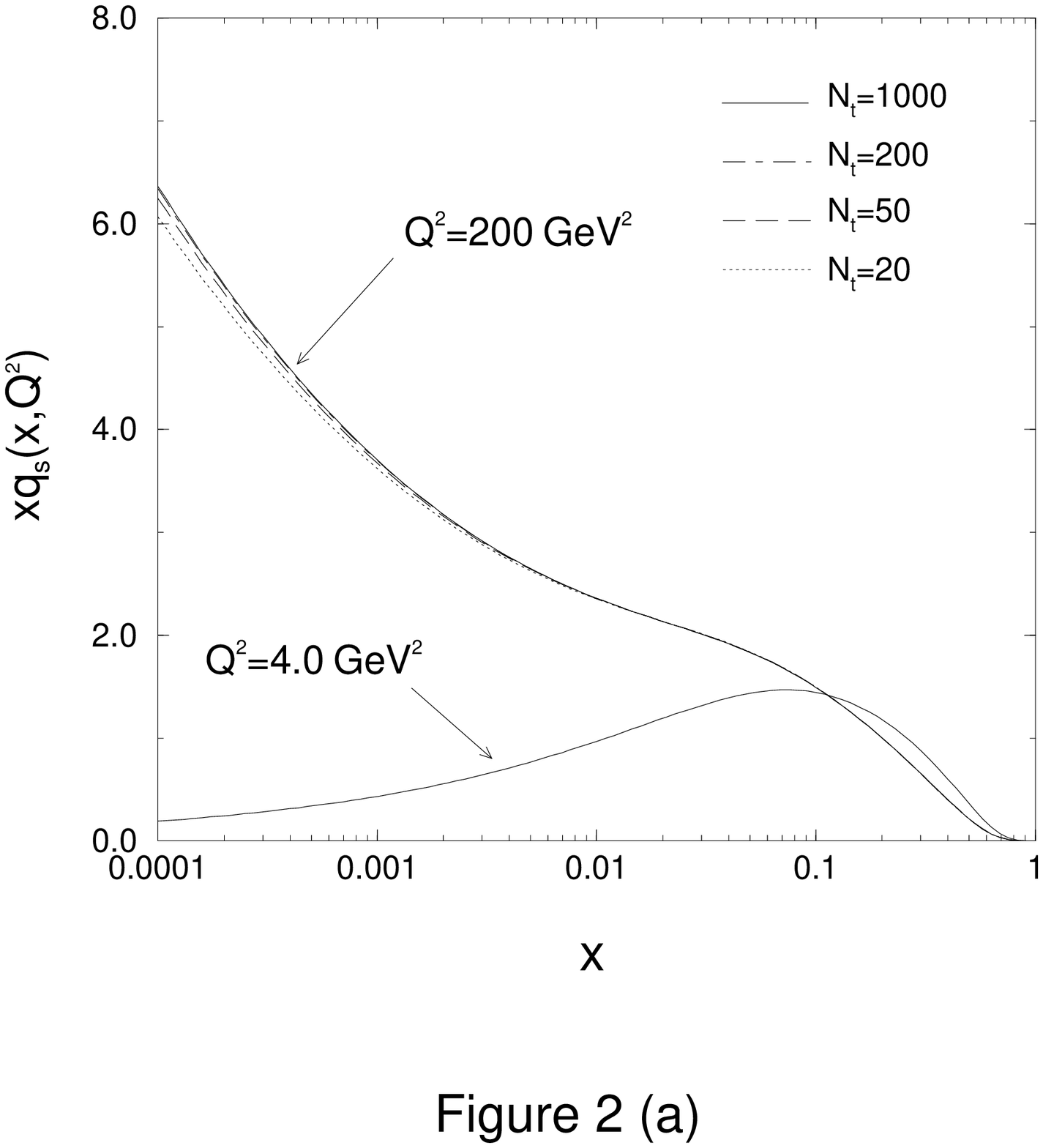}

\hspace{2.7cm}
\epsfysize=10.5cm
\epsfbox{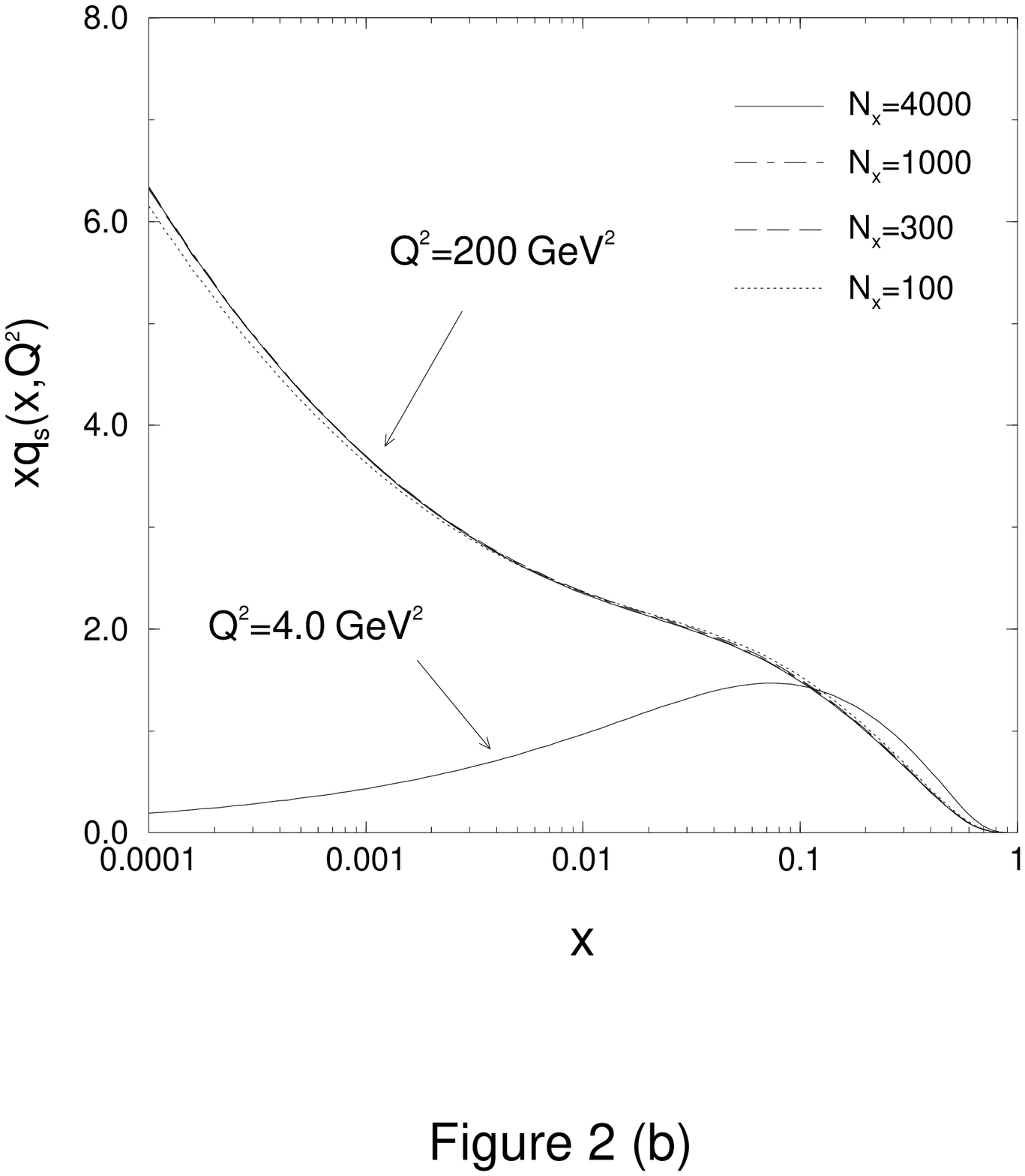}
\vfill\eject
\hspace{2.7cm}
\epsfysize=10.5cm
\epsfbox{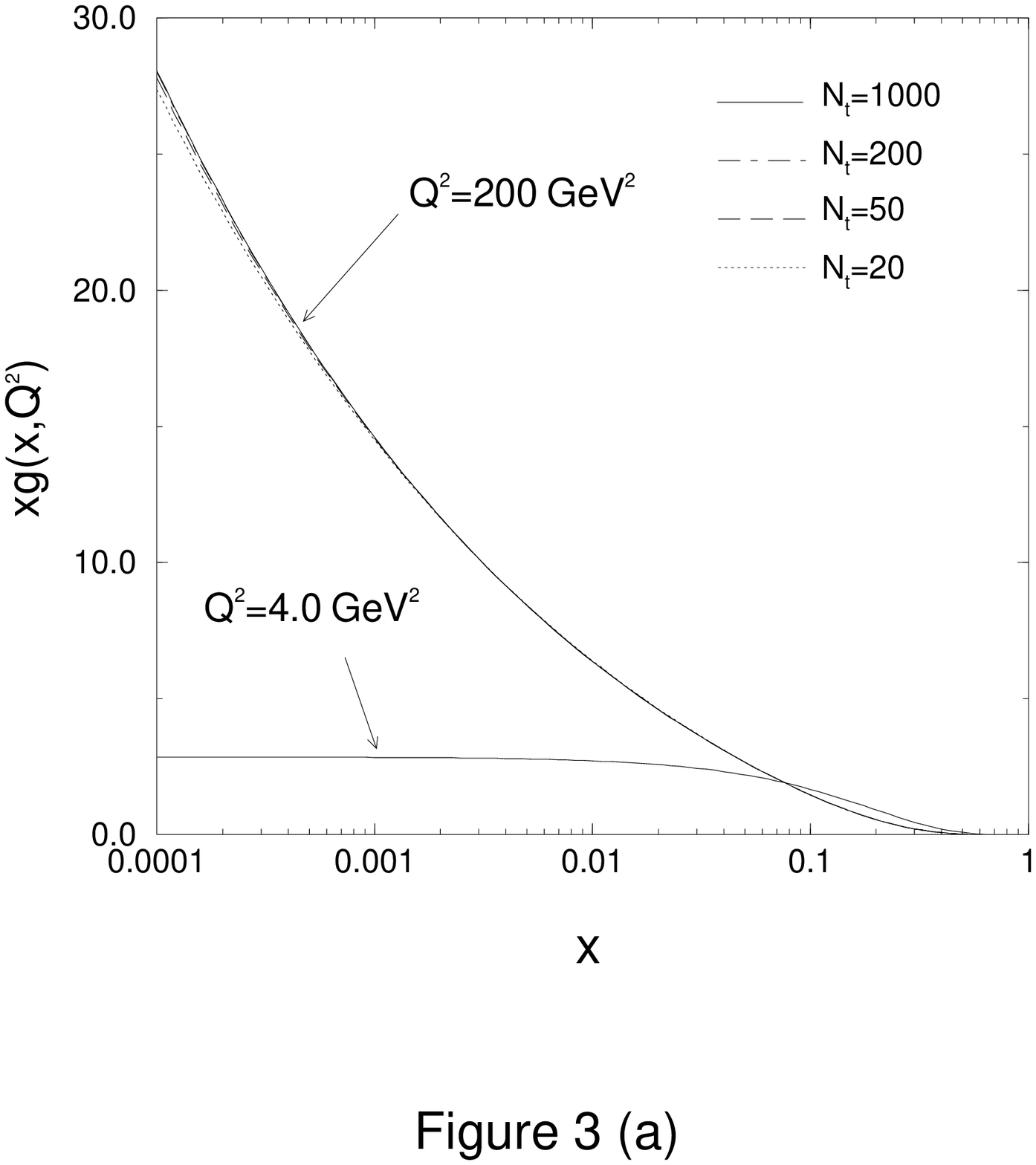}

\hspace{2.7cm}
\epsfysize=10.5cm
\epsfbox{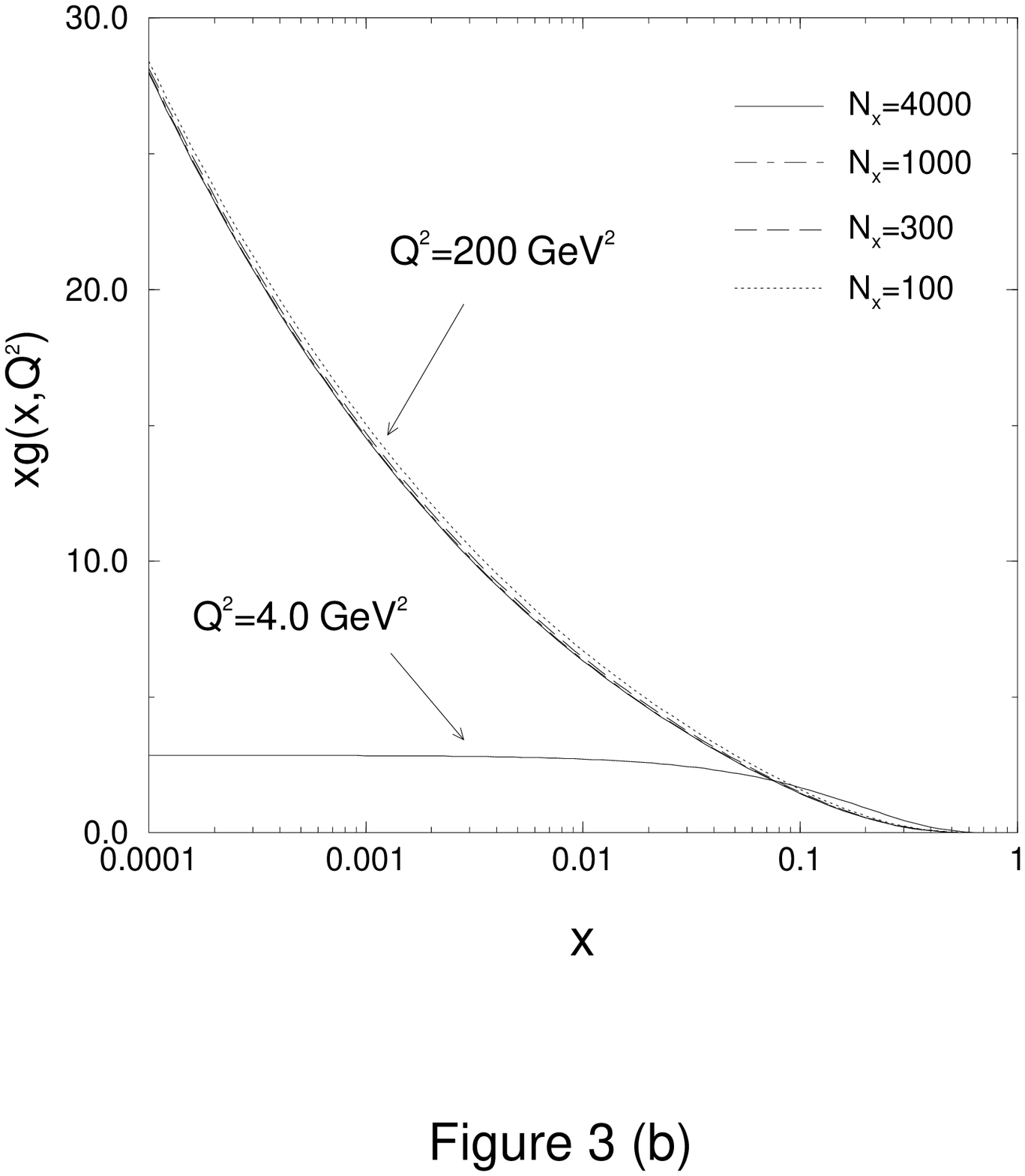}
\vfill\eject
\epsfysize=20.0cm
\epsfbox{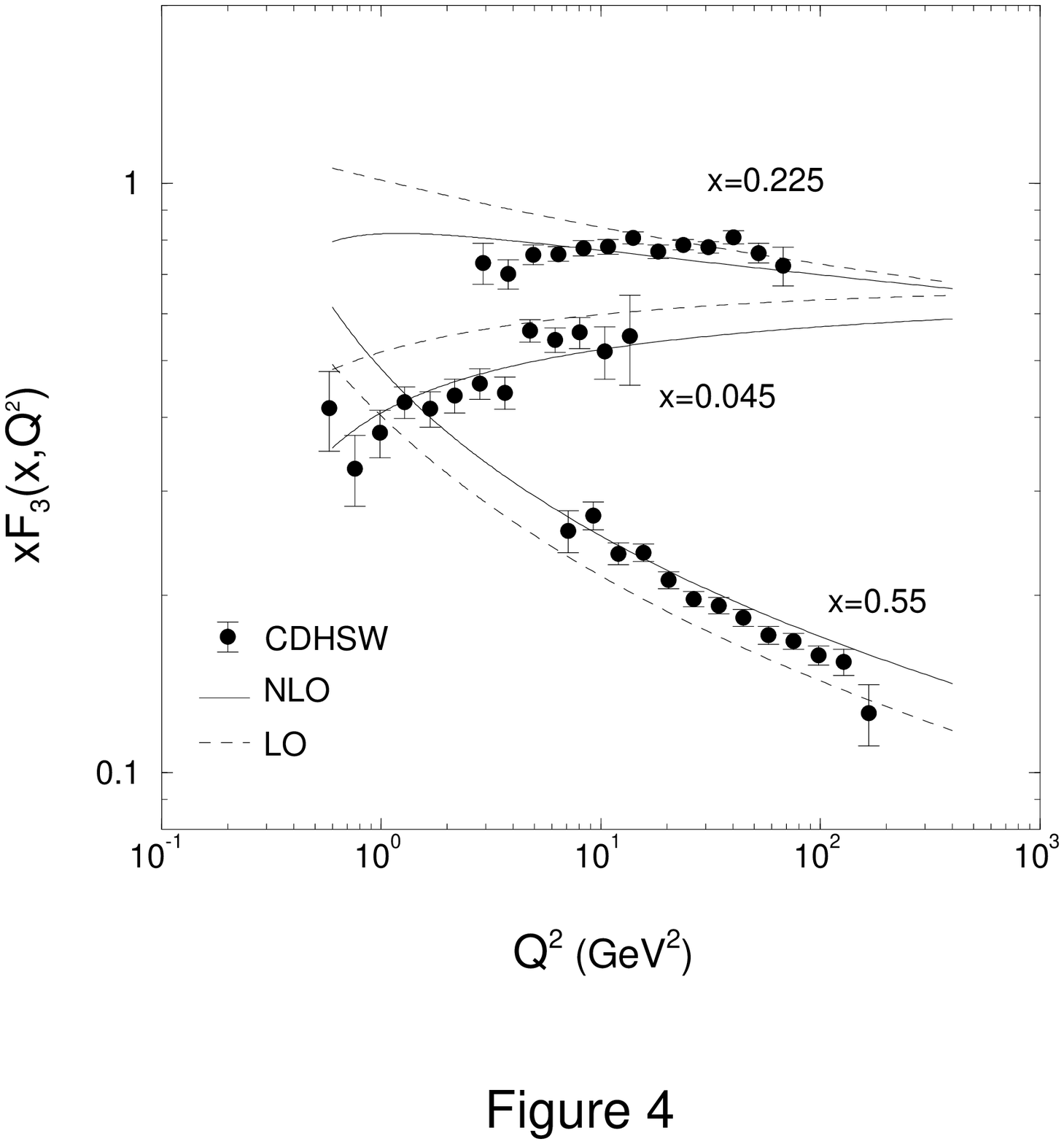}
\vfill\eject
\epsfysize=20.0cm
\epsfbox{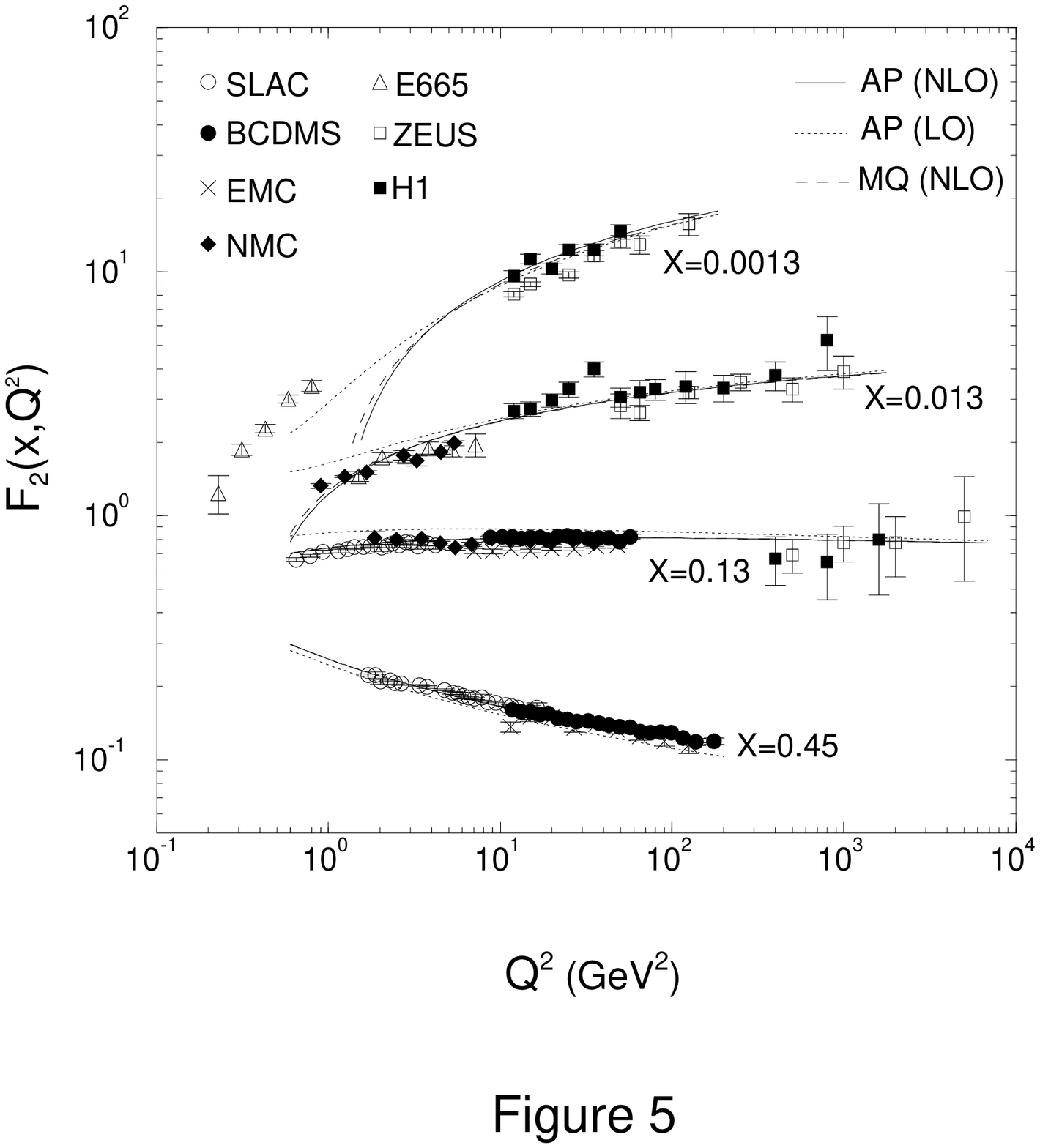}
\vfill\eject
\hspace{2.7cm}
\epsfysize=10.5cm
\epsfbox{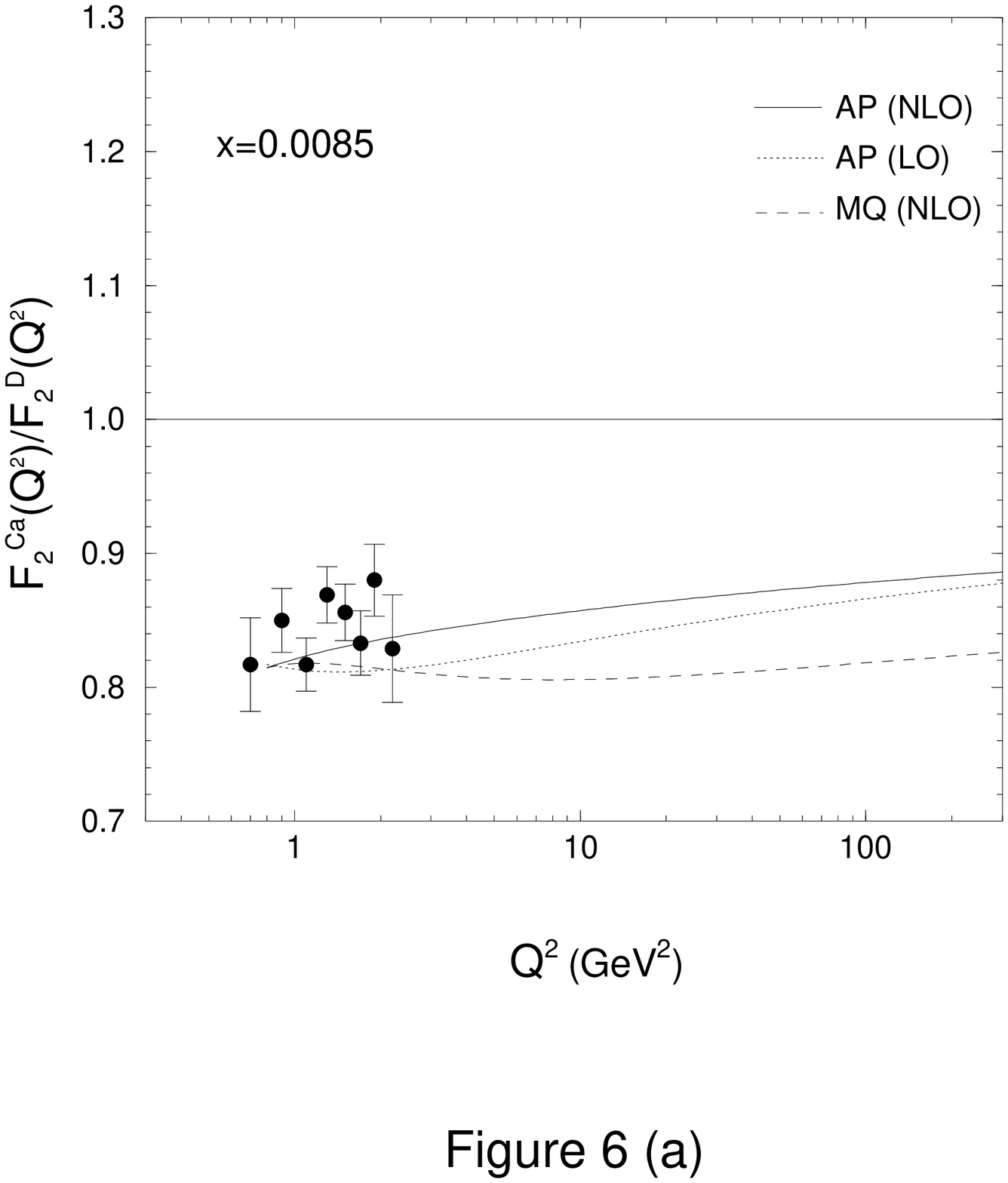}

\hspace{2.7cm}
\epsfysize=10.5cm
\epsfbox{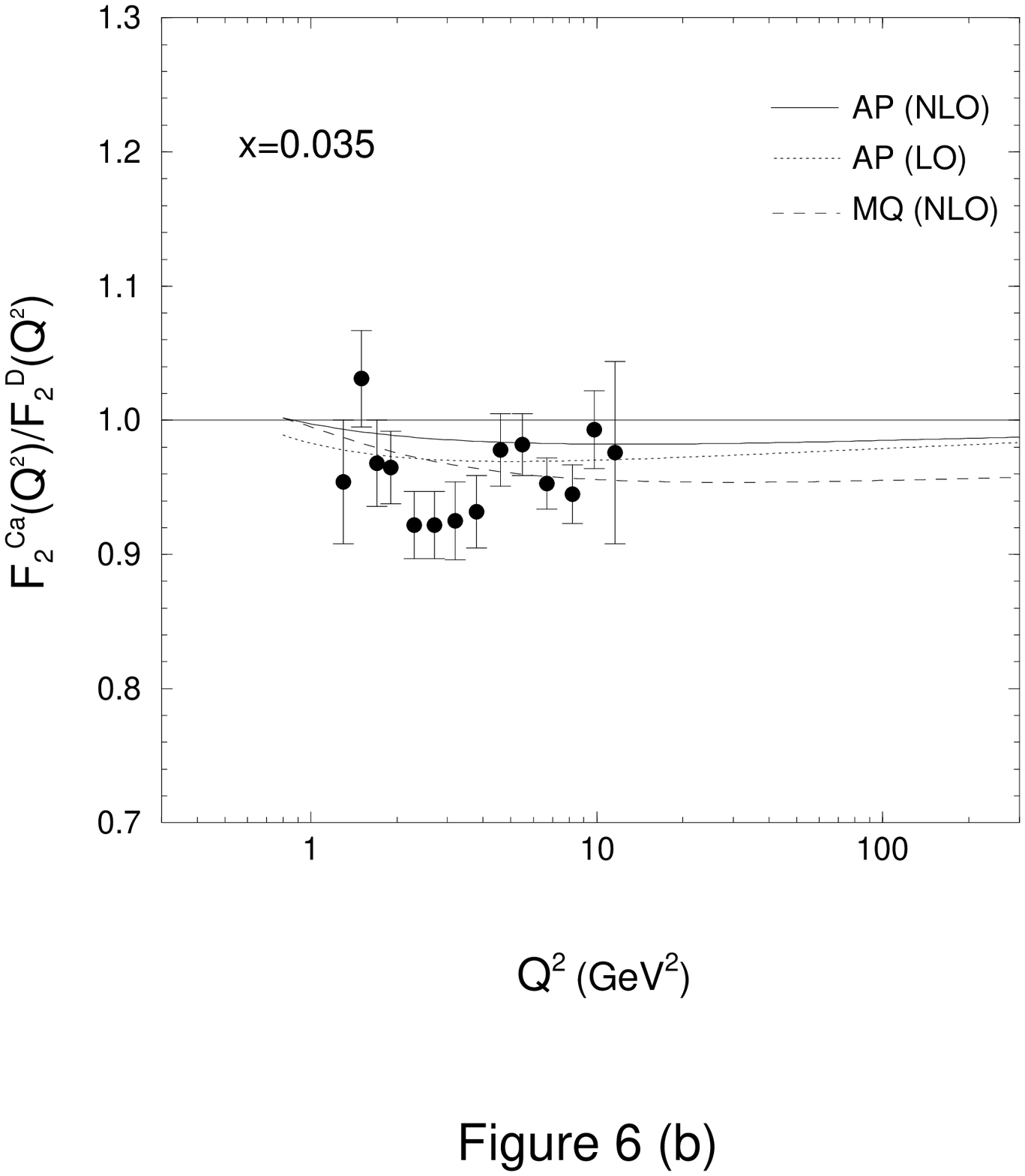}
\vfill\eject
\hspace{2.7cm}
\epsfysize=10.5cm
\epsfbox{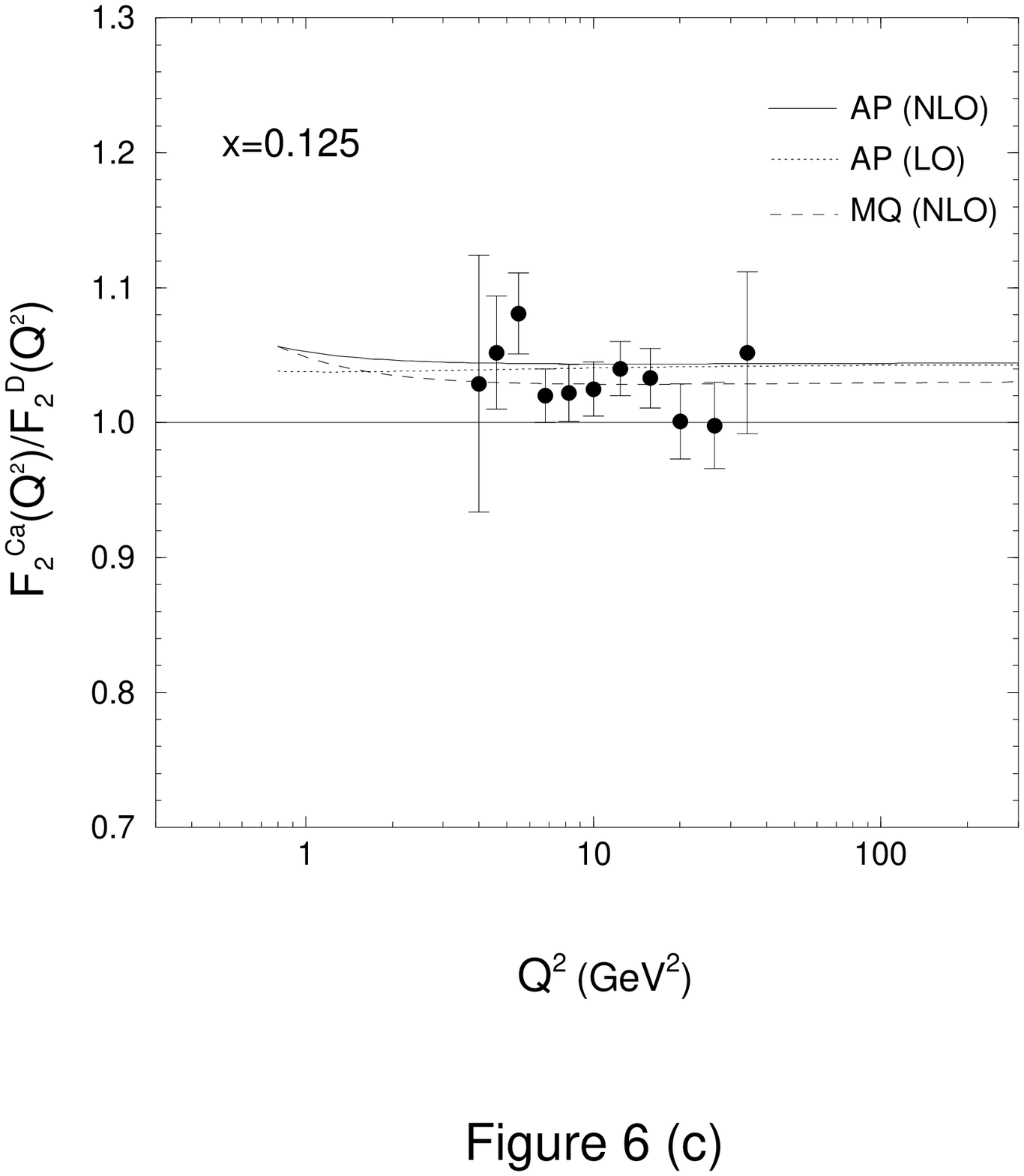}

\hspace{2.7cm}
\epsfysize=10.5cm
\epsfbox{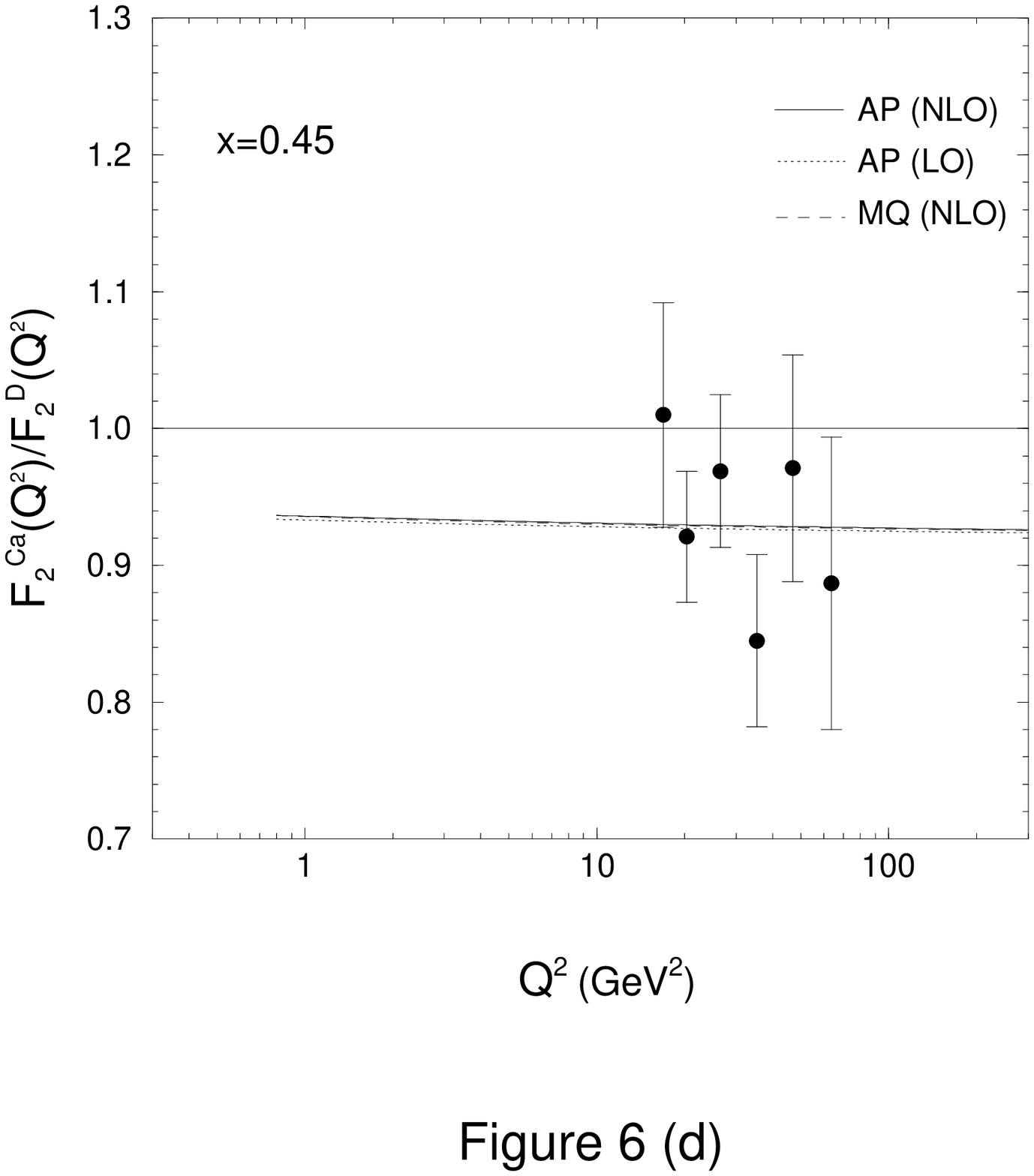}


\begin{thebibliography}{}
\bibitem{SOLVE} References to various methods are listed in the
                RK, MK, and SK paper in Ref. \cite{LAG}.
\bibitem{SLAC} SLAC data in the report SLAC-PEP-357 are taken
               from \cite{RWDATA}.
\end{thebibliography}
\end{document}